\documentclass[final,5p,times,twocolumn]{elsarticle}

\usepackage{amssymb}
\usepackage{amsmath}

\usepackage{algorithmic}
\usepackage{array}

\usepackage{textcomp}
\usepackage{stfloats}
\usepackage{url}
\usepackage{verbatim}
\usepackage{graphicx}
\usepackage{booktabs}
\usepackage{multirow}
\usepackage{rotating}
\usepackage{subfigure}
\usepackage[colorlinks=true]{hyperref}
\usepackage[table,xcdraw,dvipsnames]{xcolor}
\usepackage{lscape}

\usepackage{caption, subcaption, times, float, cuted}

\usepackage{pifont}
\newcommand{\cmark}{{\color[HTML]{34FF34} \textbf{\ding{51}}}}%
\newcommand{\xmark}{{\color[HTML]{FE0000} \textbf{\ding{55}}} }%
\newcommand{\cmarkblack}{{\color[HTML]{000000} \textbf{\ding{51}}}}%

\usepackage{balance}

\newcommand{\Rebuttal}[1]{\textcolor{black}{#1}}
\newcommand{\SecondRebuttal}[1]{\textcolor{black}{#1}}

\journal{Signal Processing and Image Communication}

\begin{document}

\begin{frontmatter}



\title{Signal Processing for Haptic Surface Modeling: a Review}



\author[1,3]{Antonio Luigi Stefani\corref{cor1}}\ead{antonioluigi.stefani@unitn.it}
\author[1,3]{Niccolò Bisagno}
\author[2]{Andrea Rosani}
\author[1,3]{Nicola Conci}
\author[1,3]{Francesco De Natale}

\affiliation[1]{organization={University of Trento},
            addressline={Via Sommarive, 9}, 
            city={Trento},
            postcode={38123}, 
            state={Trento},
            country={Italy}}
\affiliation[2]{organization={Free University of Bozen-Bolzano},
            addressline={Via Bruno Buozzi, 1}, 
            city={Bolzano},
            postcode={39100}, 
            state={Bolzano},
            country={Italy}}
\affiliation[3]{organization={CNIT - Consorzio Nazionale Interuniversitario per le Telecomunicazioni},
            addressline={Via Sommarive, 14}, 
            city={Trento},
            postcode={38123}, 
            state={Trento},
            country={Italy}}

\cortext[cor1]{Corresponding author}


\begin{abstract}
Haptic feedback has been integrated into Virtual and Augmented Reality, complementing acoustic and visual information and contributing to an all-round immersive experience in multiple fields, spanning from the medical domain to entertainment and gaming. Haptic technologies involve complex cross-disciplinary research that encompasses sensing, data representation, interactive rendering, perception, and quality of experience.
The standard processing pipeline, consists of (I) sensing physical features in the real world using a transducer, (II) modeling and storing the collected information in some digital format, (III) communicating the information, and finally, (IV) rendering the haptic information through appropriate devices, thus producing a user experience (V) perceptually close to the original physical world. 
Among these areas, sensing, rendering and perception have been deeply investigated and are the subject of different comprehensive surveys available in the literature. Differently, research dealing with haptic surface modeling and data representation still lacks a comprehensive dissection.
In this work, we aim at providing an overview on modeling and representation of haptic surfaces from a signal processing perspective, covering the aspects that lie in between haptic information acquisition on one side and rendering and perception on the other side.
We analyze, categorize, and compare research papers that address the haptic surface modeling and data representation, pointing out existing gaps and possible research directions.
\end{abstract}





\end{frontmatter}



\section{Introduction}
\label{sec:intro}




\begin{figure*}[!h]
    \centering
    \includegraphics[width=1\linewidth]{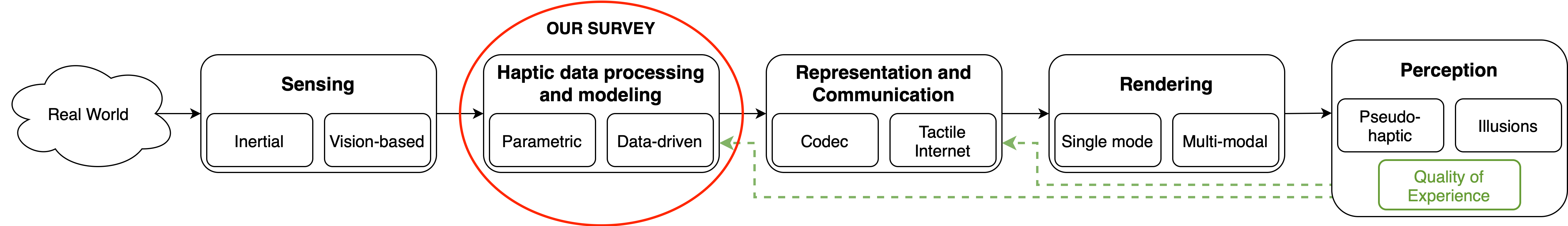}
    \caption{The haptic processing pipeline consists of 5 main steps: sensors to acquire the data, processing and modeling to store acquired data, communication to transmit information associated with the collected signals, haptic devices to render the haptic feedback to the user, and perception that studies the factor that influence the haptic experience of the user.}
    \label{fig:haptic_pipeline}
\end{figure*}

The primary objective of extended-Reality (XR), in all its possible variants (virtual-, augmented-, mixed-) is to either faithfully replicate real-world experiences or enhance them \cite{xi2023challenges,alizadehsalehi2020bim}, enriching the real-world information by overlaying additional virtual contents\cite{keshner2004virtual,jungherr2022extended}. To effectively complement and/or replace their actual real-world counterparts, virtual experiences must be designed to engage as much as possible the various human senses \cite{kral2022digital}.

For decades, visual and acoustic stimuli have been the focal points of XR solutions, due to the large availability of reliable and affordable technologies. For example, current AR/VR headsets are able to render realistic immersive environments with synchronized and spatialized audio-visual information. However, whenever the application requires the interaction with virtual objects, the need of engaging user's touch emerges, being the only way to provide a realistic perception of the contact with objects. Notwithstanding their importance in this respect, haptic technologies have remained relatively unexplored, lagging behind the rapid development of spatial audio and high-definition images, and with very little impact on the market, especially at a consumer level.

To behave in a more natural way when interacting with a virtual environment, users' sense of touch should also be stimulated, allowing them to grasp a virtual object, and feel the roughness of its surface.
To do so, researchers focused on how to generate haptic feedback as close as possible to the real world. This implies a number of operations to be performed, which span different technological domains. Accordingly, we propose in Fig. \ref{fig:haptic_pipeline} a breakdown of the haptic rendering process that defines in a clear and interpretable manner the steps that are required to effectively achieve such feedback. 

In particular, the overall process can be split into five main blocks, namely: (I) sensing, (II) haptic data processing and modeling, (III) representation and communication, (IV) rendering, and (V) perception. Sensors, such as inertial or vision-based sensors, are employed to scan objects and capture their haptic properties, such as stiffness, roughness, and slipperiness. Haptic data processing and modeling focuses on the encoding of the collected physical information to some digital format (images, haptic maps, spectrograms) and its processing mechanisms. This is also referred to as Haptic Surface Modeling. The block of representation and communication of haptic data studies how to efficiently compress, encode and transmit such information. Rendering is performed by haptic devices (e.g., haptic gloves, haptic displays), which are able to generate an appropriate stimulus starting from the captured model. Finally, user's feedback (perception) and the Quality of Experience (QoE) are studied to understand how close the haptic interaction is to the real-world touch feeling, making use of pseudo-haptic and haptic illusions techniques.

In Tab. \ref{tab:surveys}, we list the surveys currently available in the literature, covering the different areas of the haptic pipeline. As can be seen, the focus of such works is either on sensors, or on the later stages of rendering and perception. A few works are also classified as 'general' (dealing with cross-disciplinary research in collateral areas, such as psychology, perception, or engineering), and application-oriented (dealing with real-world setups and systems that make use of haptics). None of the listed works, however, explores in detail the available literature in the area of Haptic Surface Modeling.

\begin{table}[!h]
\centering
\resizebox{\linewidth}{!}{%
\begin{tabular}{@{}ccc@{}}
\textbf{Topic}                                                                                  & \textbf{Ref.}                                                                                                                                                                              & \textbf{Description}                                                                                                                                          \\ \hline
\textbf{General}                                                                                & \cite{biswas2021haptic, culbertson2018haptics, fleury2020survey, laycock2007survey, salisbury2004haptic, varalakshmi2012haptics}                                                           & \begin{tabular}[c]{@{}c@{}}General topics with convergent research \\ on haptic perception, mechanics, \\ electronics, and material technologies\end{tabular} \\ \hline \hline
\textbf{(I) Sensors}                                                                            & \cite{chen2018tactile, kappassov2015tactile, yamaguchi2019recent, zhang2022hardware}                                                                                                       & \begin{tabular}[c]{@{}c@{}}Hardware-related topics to collect \\ haptic data, such as vision-based and \\ inertial sensors\end{tabular}                       \\ \hline
\textbf{\begin{tabular}[c]{@{}c@{}}(II) Haptic data \\ processing and \\ modeling\end{tabular}} & \textbf{Our}                                                                                                                                                                               & \begin{tabular}[c]{@{}c@{}}Focusing on Haptic Surface Modeling, \\ thus how the haptic data can be stored, \\ processed and analysed\end{tabular}             \\ \hline
\textbf{\begin{tabular}[c]{@{}c@{}}(III) Representation\\ and communication\end{tabular}}       & \cite{promwongsa2020comprehensive,minopoulos2019survey,kokkonis2012survey,awais2023towards,antonakoglou2018toward}                                                                         & \begin{tabular}[c]{@{}c@{}}Focusing on allowing \\ the transmission of haptic information\\ over a network (Tactile Internet)\end{tabular}                    \\ \hline
\textbf{\begin{tabular}[c]{@{}c@{}}(IV) Haptic \\ devices\end{tabular}}                         & \cite{adilkhanov2022haptic, basdogan2020review, breitschaft2022haptic, choi2012vibrotactile, costes2020towards, mercado2021haptics, tan2020methodology, wang2019multimodal, wee2021haptic} & \begin{tabular}[c]{@{}c@{}}Multimodal and single mode devices to \\ convey the haptic sensation to the user\end{tabular}                                      \\ \hline
\textbf{(V) Perception}                                                                         & \cite{kurzweg2024survey, lecuyer2009simulating, ujitoko2021survey}                                                                                                                         & \begin{tabular}[c]{@{}c@{}}Focusing on how realistic is the tactile\\ sensation perceived by the user\end{tabular}                                            \\ \hline \hline
\textbf{Applications}                                                                           & \cite{coles2010role, escobar2016review, gaffary2018use, hamza2019haptic, imran2021significance, luo2017robotic, petermeijer2015effect, saint2021survey}                                    & \begin{tabular}[c]{@{}c@{}}Various applications of the haptic\\ rendering process\end{tabular}                                                               
\end{tabular}%
}
\caption{Available surveys in the domain of haptics focus on different steps of the haptic rendering process, but none of them zeros in on the data representation for haptic signals.}
\label{tab:surveys}
\end{table}

\Rebuttal{In this work we focus on the problem of associating the visual aspect of a textured surface (its image) to the relevant haptic information and we review the signal processing methodologies that allow to extract and model such information. This is a fundamental aspect for the image/video communication systems of the future and for the construction of truly immersive AR/VR systems \cite{chen2024comprehensive}. Indeed, although some attempts have been proposed in the past \cite{kim2013construction}, at present there are no standards for managing haptic information in multimedia content streaming. We believe that the progress in this field builds upon consolidated and universally accepted methodologies to associate image and haptic information, and signal processing is certainly a key enabling technology in this context.}

In this survey, we aim at reviewing such missing key aspect, 
i.e., the representation of real-world haptic features in the digital domain. We provide a comprehensive review on the representation of haptic surfaces from an image and signal processing perspective. We examine, which data are mostly used in today's applications, looking at the existing datasets, and we indicate how they are being adopted by the research community. Drawing a parallelism with the research in the visual and acoustic domains, we highlight the shortcoming of the current haptic processing pipelines, pointing out possible research directions. \Rebuttal{Although the sense of touch can be experienced across the entire human body \cite{ke2022propelwalker,ushiyama2023feetthrough}, this survey focuses primarily on hand interactions, which are prevalent in the available literature and commercial devices.}

The paper is organized as follows. In Section \ref{sec:defintions}, we introduce the definitions that will be used throughout the paper. Section \ref{sec:processing_pipeline} delineates the current processing pipelines for the visual and acoustic signals and compares them to the haptic one, highlighting differences and similarities. In Section \ref{sec:haptic_surface_modeling}, we present the available literature on Haptic Surface Modeling and processing. Section \ref{sec:datasets} showcases available datasets for haptic processing research, while in Section \ref{sec:applications} we highlight current research tasks, including baselines and metrics. Section \ref{sec:conclusions} concludes the paper, providing a critical discussion and highlighting future trends and open research issues.




\section{The Haptic Processing Pipeline: Definitions}
\label{sec:defintions}


According to the representation in Fig. \ref{fig:haptic_pipeline}, when dealing with haptic technologies we can identify five main building blocks, which contribute to the definition of the overall pipeline, namely: sensing, processing and modeling, representation and communication, rendering, and perception. In this paragraph we briefly describe each of these elements, introducing definitions and concepts that will be used throughout the paper, and providing links to the relevant surveys for each specific topic.

\subsection{Sensing}
\label{sec:sensors}

Haptic sensors aim at mimicking the receptors that are present in our fingers and hands, to extract essential tactile information. \Rebuttal{Although many types of sensors exist to collect haptic data, the two types of sensors majorly employed to model haptic properties are inertial and vision-based}. Inertial sensors adopt various technologies (resistive, capacitive, fiber-based) to detect the roughness of the surface through the contact with it. They produce one-dimensional signals that capture the forces of interaction between sensor and object, which convey information about surface properties. Inertial sensors have been largely employed in robotic applications \cite{kappassov2015tactile,chen2018tactile}, where tactile sensing refers to the point of contact between robot fingers and objects.

Vision-based sensors \cite{yamaguchi2019recent, zhang2022hardware} have recently emerged as the most simple and affordable way to capture tactile information at high resolution from images or video. The use of images and video appears to be very intuitive to tackle the problem. This is generally implemented using a RGB camera recording the deformation of a membrane upon its contact with a surface, thus generating a haptic map.





\subsection{Haptic data processing and modeling}
\label{sec:data_representation}
Haptic data processing and modeling, also called Haptic Surface Modeling, investigates how the sensed haptic signals are handled and stored in digital form. Unlike other sensory domains such as visual \cite{susstrunk1999standard} and acoustics \cite{fletcher2012physics}, where standardized representations are commonly available, the representation of tactile stimuli remains an area without established standards.
In the early stages, haptic surface modeling relied on parametric functions to model the surface shape \cite{basdogan2002haptic}. However, this approach either introduced a high amount of noise and approximation in the model or relied on too complex mathematical expressions. Consequently, researchers have opted for a paradigm change, moving towards data-driven approaches \cite{hover2008data}, where the object's haptic information is collected as a one-dimensional signal, 
associated to classical representations such as spectrograms. The natural evolution of these initial approaches was then to turn to 2D signals, such as images and videos as well as haptic maps, which provide a much richer information that can be approached 
through well-studied computer vision techniques.

In this survey, we focus on these aspects, investigating the different options in terms of processing and highlighting advantages and shortcomings for each of them.

\subsection{Representation and communication}
\label{sec:representation_communication}

\Rebuttal{The representation of haptic data is still underexplored, complicating the integration of haptic stimuli within virtual environments. Although beyond the scope of this survey, we provide a brief overview of the literature addressing the representation of haptic information. In \cite{wakita2008texturebased}, the authors define a method to encode haptic information into 3 maps, each representing a specific haptic stimulus. These maps include height maps, friction maps, and stiffness maps \cite{wakita2008texturebased,wakita2010friction}. Similarly to this approach, in \cite{kamuro2012haptic}, the authors developed an editor that uses four layers to define the haptic stimuli. In \cite{kim2011haptic}, the authors extract haptic information from visual images and represent it using depth, stiffness, and viscosity maps.
\\
In \cite{costes2018haptic}, the authors draw a more complete overview of the previous methods. Furthermore, they define 10 maps correlating haptic characteristics (e.g., compliance, geometry) with perceptual modalities (e.g., kinesthetic, thermal).}

Even more than its visual and audio counterparts, the haptic signal needs to rely on low latency communication to develop real-time interactive systems \cite{awais2023towards}. Potential applications include tele-surgery, industrial processes automation and augmented reality for live music concerts \cite{promwongsa2020comprehensive}. To enable such applications, researchers have focused on building the so-called Tactile Internet \cite{promwongsa2020comprehensive,steinbach2018haptic,awais2023towards}, elaborating on suitable technologies and communication protocols in the different layers of an architecture, such as the information representation using suitable codecs \cite{steinbach2018haptic}, as well as network design and customer's haptic devices.

\subsection{Haptic devices}
\label{sec:haptic_devices}

Haptic devices aim at providing the tactile experience to the user, effectively translating digital data into physical stimuli that hit the receptors of human haptic channels \cite{wang2019multimodal}. The haptic stimulus can be delivered using single mode or multi-modal devices. Among single mode devices, surface haptic displays \cite{costes2020towards} consist of an active physical planar device that deforms so as to reproduce a surface model, usually sensed by the bare finger. Differently, multimodal haptic devices, like hand-held (joystick-like, unconstrained devices), wearables (usually gloves) and encountered/grounded (surfaces/devices for the user to touch "on-demand") \cite{wang2019multimodal,basdogan2020review,adilkhanov2022haptic,wee2021haptic} aim at exploiting different stimuli, such as forces, vibrations and thermal, to convey the haptic sensation to the user.



\subsection{Perception}
\label{sec:perception}

Human haptic perception encompasses a range of sensory modalities, which can act externally in the form of cutaneous sensing, and internally as proprioception and kinesthesia \cite{kurzweg2024survey}. While single mode haptic devices aim at reproducing pseudo-haptic feedback \cite{lecuyer2009simulating,ujitoko2021survey}, they struggle to fulfill the whole range of human haptic perception. Multi-modal devices could expand the spectrum of haptic feedback, reproducing the sensations produced by touching an object rather than its physical features, thus creating the so-called haptic illusions \cite{kurzweg2024survey}. In human-centered applications, Quality of Experience (QoE) plays a crucial role in influencing the design and overall success of the whole pipeline. However, QoE in haptics is usually evaluated through subjective tests with the user-in-the-loop \cite{steinbach2018haptic}. The desirable evaluation of QoE through objective testing aims at defining algorithmic models of human perception to calculate meaningful perceptual metrics \cite{steinbach2018haptic,culbertson2018haptics}.
Other than human-based applications, perception is also key in robotics, for instance when dealing with object grasping \cite{luo2017robotic,li2020review}. In such scenarios, understanding objects' properties plays a fundamental role to modulate the applied force in the grasping action and to minimize failures.




\section{Real world objects representation in the physical and the perceptual spaces}
\label{sec:processing_pipeline}

Real-world objects are defined by their physical characteristics, such as color, shape, roughness, and produced sound. Our perception of an object depends on these characteristics as well as on a number of subjective factors, such as sensory sensitivity and cognitive processing. To develop a digital model for virtual interactions, both physical and perceptual characteristics must be considered.

\begin{figure*}
    \centering
    \includegraphics[width=0.90\textwidth]{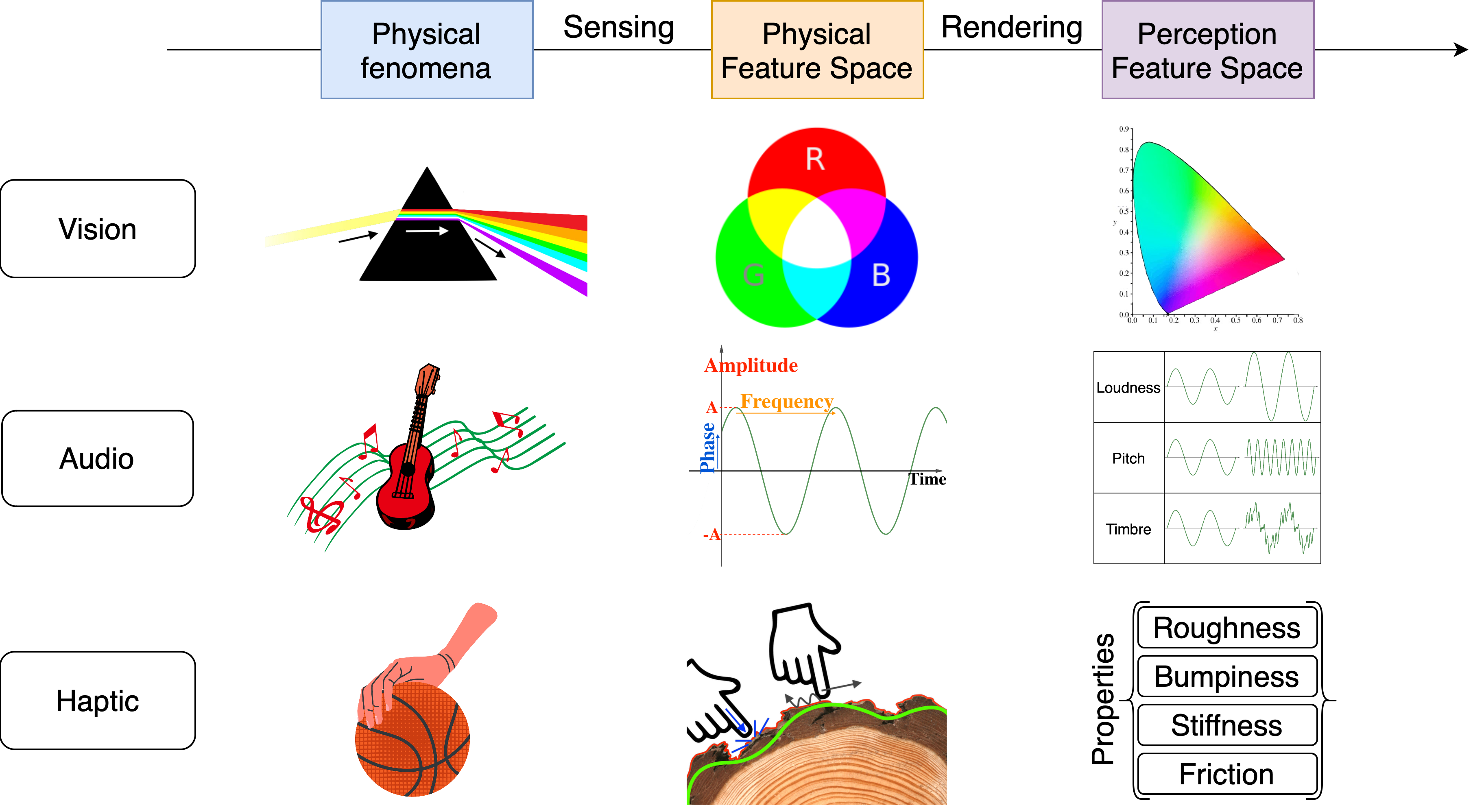}
    \caption{The classic signal processing pipeline (top row) aims at mapping a physical phenomenon to a physical feature space such that a similar experience of the phenomenon can be delivered to the user in the perception feature space (bottom row). While the pipeline is well-established in both the visual and acoustic domains, it still lacks a common standard in the haptic domain. \Rebuttal{In the example in bottom row illustrating the section of a tree,} the \textbf{\textcolor{red}{red} }line is the roughness, the \textbf{\textcolor{green}{green}} line is the bumpiness, the \textbf{\textcolor{blue}{blue}} line represents the stiffness, and the \textbf{\textcolor{darkgray}{dark gray}} line is the friction.}
    \label{fig:representation_pipeline}
\end{figure*}

Fig. \ref{fig:representation_pipeline} illustrates the modeling pipeline, which connects physical phenomena sensed by specific sensors, the physical feature space that defines the object's characteristics, and the perception feature space that identifies its properties as they are perceived by humans. While physical and perceptual spaces are well-established for images and audio signals, haptic signals still lack a standard representation able to characterize and replicate the relevant properties \cite{degraen2021capturing}.

In the following paragraphs we highlight the existing definitions of the physical and perceptual spaces for both visual and acoustic domains, trying to draw a parallelism with the haptic domain, and highlighting gaps and possible research directions.




\subsection{Physical feature space}
The physical feature space contains the characteristics to be captured in order to represent the physical phenomenon of interest. The role of the sensor is to capture one or more physical characteristics of the phenomenon in a given domain. As an example, cameras provide information in the visual domain by capturing wavelengths in the visible (or infrared) spectrum, while microphones record sounds in terms of frequency, amplitude and phase. When dealing with haptics, roughness and stiffness are among the most common features to be sensed. To better explain haptic sensing, in the next paragraphs, we will draw an analogy between the haptic domain and the more familiar visual and audio domains.


\paragraph{Visual domain} \Rebuttal{The RGB color model \cite{susstrunk1999standard} is the most widespread model to describe the physical lights properties in the visual domain \cite{susstrunk1999standard, ibraheem2012understanding}. This model identifies three main wavelengths perceived by our eye, which are then combined to synthesize all the other colors of the visible spectrum. To convert colors to a digital representation, a digital RGB camera measures the quantity of photons hitting the sensor. Then, the sensors converts the photons quantity into an electric current which voltage value is quantized to obtain to the digital representation of the original color.} 

\paragraph{Acoustic domain} \Rebuttal{Acoustic waves are the signal generated by the vibrations of a source. Their physical characteristics are usually described by 3 quantities known as frequency, amplitude and phase \cite{fletcher2012physics}.}
\Rebuttal{More specifically, the frequency describes the pitch of the sound, the amplitude describes its strength, and the phase describes the position of a sound wave relative to a reference point in its cycle \cite{fletcher2012physics, muller2015fundamentals}. Similarly to the visual domain, the acoustic waves are first collected by a microphone equipped with a membrane, whose vibrations are converted to an electric signal and then digitized.}

\paragraph{Haptic domain} The properties describing the haptic qualities of an object in the physical space and a standard model encapsulating them did not reach a consensus among researchers yet. Nevertheless, as will be seen in the next paragraphs, researchers agree on the inclusion of some common macro categories as, for example, the concepts of roughness and friction. \Rebuttal{In fact, researchers have focused more on the perception feature space, living behind the studies aiming at characterizing which are the physical properties that can be used to describe the haptic sensations.} 

The first attempt to define a haptic physical space and to correlate the physical properties of a material with their perception is described in \cite{chen2009exploring}. In this work, the authors identified four physical properties that characterize a surface, namely: roughness, compliance, friction, and cooling rate. According to the authors, roughness measures the  variations of the height of the surface, compliance measures its softness, friction measures the sliding resistance of the surface, and cooling rate measures the variations of temperature across it.

More recently, in \cite{hassan2019authoring, hassan2023establishing} the authors proposed a new way to characterize objects according to a set of haptic properties that can be considered orthogonal in what they called Haptic Attribute Space (HAS).
In particular, the attributes that describe the four-dimensional HAS space are defined as follows: \textbf{roughness} (or micro-roughness) corresponds to the finer spaced irregularities of the surface texture that usually result from the inherent action of the production process or material condition \cite{american2003surface}; \textbf{bumpiness} (also called macro-roughness or waviness) is associated to the macro-behavior of the surface texture, such as ridges, bumps and valleys; 
\textbf{stiffness} (or elasticity), it is the term used to describe the force needed to achieve a certain deformation of a structure \cite{baumgart2000stiffness}; \textbf{friction} (or slipperiness) refers to the resistance that one surface or object encounters when moving over another. In other words, it is the force that opposes the relative motion or tendency of such motion of two surfaces in contact \cite{bowden2001friction}. In \Rebuttal{Fig.} \ref{fig:representation_pipeline}, a visual example is depicted of the above properties.



Despite some attempts to define a standardized physical feature space for haptic data, a universally accepted representation has not emerged yet. Furthermore, the physical space is often conflated with the perception space, and this frequently generates confusion and misunderstandings  when describing applications related to sensing and actuating surfaces that recreate haptic perceptions.

\subsection{Perception feature space}

The perception feature space aims at describing the properties that enable us to sense and interpret information from the surrounding world. These properties have been the subject of study for decades, as they inform the design of sensors and actuators. Also in this case, to better highlight the underlying concepts, we propose a simple parallel with other known perceptual spaces, in the visual and audio domains.


\paragraph{Visual domain} \Rebuttal{The CIELAB standard \cite{standard2007colorimetry, cheung2012uniform}, an evolution of the initially proposed CIE XYZ 1931 \cite{smith1931colorimetric}, aims at quantitatively describing the relation between the wavelength distributions in the electromagnetic visible spectrum and the physiologically perceived colors in human vision \cite{stiles1959npl}. The CIELAB color space is built starting from the human vision and, differently from the RGB space, it is device-independent (e.g., the same RGB triplet can look different on separate devices, and it is thus mapped differently in the CIELAB space), \SecondRebuttal{although almost all devices currently rely on automatic gamut adaptation to the device (like sRGB).} The relation between RGB and CIELAB spaces has been defined in \cite{connolly1997study}. Subsequent works \cite{fdhal2009color,ibraheem2012understanding} have continued the investigation on the relationship between the physical and perception spaces as well as the classification of the color spaces and how they are built.}

\paragraph{Acoustic domain} \Rebuttal{The perception space in the acoustic domain is characterised by 3 main features: Pitch, Loudness, and Timbre \cite{muller2015fundamentals, moore2012introduction}. Differently from the acoustic physical space, these features have been already used to define a three-dimensional space, where sound waves can be considered as vectors. However, many studies correlated these properties to the physical space dimensions, namely: Frequency, Amplitude and Phase. The Pitch of a sound is closely related to the frequency notion, but it refers to our perception of the fundamental frequency \cite{zwicker2013psychoacoustics, muller2015fundamentals}. The Loudness dimension refers to the sound intensity \cite{fletcher1933loudness, moore2012introduction} and its value varies on a scale from quiet to loud \cite{muller2015fundamentals}. The Timbre \cite{sethares2005tuning} is the characteristic allowing us to recognize the tone color of the sound we are hearing.}

\paragraph{Haptic domain}





Also in this case, differently from visual and acoustic stimuli, haptic stimulus is quite complex to describe from a perceptual viewpoint. In fact, understanding which are the surface properties within the physical space that influence our perception when interacting with an object is still an open problem \cite{costes2020towards}. Therefore, many studies have been conducted to identify an orthogonal perceptual feature space that optimizes the representation of the properties to transmit to our skin, allowing us to recognize and categorize the surfaces.


In early studies, such as \cite{holliins1993perceptual}, the authors suggested a possible two-dimensional space that describes the haptic perception in terms of roughness and hardness. Subsequently, slipperiness was added as a third dimension \cite{hollins2000individual}. In \cite{tiest2010tactual}, which has become the most common representation, the haptic features that compose the perceptual space were extended to four, namely roughness, compliance, coldness and friction. However, in \cite{okamoto2012psychophysical}, the authors further refined this model, proposing a 5-dimensional haptic perception space that included macro-roughness, micro-roughness, coldness, hardness, and friction. Other studies \cite{baumgartner2013visual, metzger2022unsupervised} tried to apply PCA on a big set of properties classified by users, to identify the components that carry more information, in an attempt to minimise the dimensional space.
In the opposite direction, other studies proposed to further enlarge the dimensional space. In \cite{sakamoto2017exploring}, for instance, two new elements called "volume" and "naturalness", were added to the perceptual space. According to the authors, "volume" describes a shape property of the surface and "naturalness" measures how much a surface is close to the natural world (as opposed to the artificial world). Another study analyzed the neural cortex response of primates and humans when scanning a surface \cite{lieber2019high}. The authors claim that the perceptual representation of a texture is high-dimensional and that the four properties usually adopted are not capable of conveying all the descriptors that are needed to uniquely describe a surface.




Notwithstanding all the recent research efforts, the four-dimensional model identified in 2010 \cite{tiest2010tactual} remains the one the research community still agrees more. Even when agreeing on these dimensions, however, it is still unclear which features carry more perceptual information (e.g., the L component in the CIELAB model conveys more information to the human eye than the other two), and how to model the relationship among features (e.g., in CIELAB, the non-linear relation between its components try to mimic the non-linear response of the human visual system).


\subsection{Linking physical and perceptual feature spaces in haptics}
While the haptic perception feature space has received broader attention from researchers, it is still unclear how to establish a direct link between the physical feature space and the perceptual feature space \cite{lu2022preference, awan2023predicting}. A first attempt to define this mapping function was made in \cite{awan2023predicting}, where the authors worked at the mapping between textures acquired by a sensor and relevant features, such as roughness and bumpiness, as perceived by the users involved in their experiment.

In short, despite the available literature, we are still far from establishing a universally accepted link between the two spaces (physical and perceptual), which remains an open problem. 

\subsection{About the temperature}


Temperature is a fundamental dimension of the haptic space, significantly affecting our exploration of the world and our interaction with it \Rebuttal{\cite{chen2009exploring,tiest2010tactual,okamoto2012psychophysical,nellis2008heat}}. \SecondRebuttal{Temperature works differently from other haptic properties, because} the sensors required to measure it (both from human and machine viewpoints) are different and complementary to those needed for other tactile properties \cite{okamoto2012psychophysical}.

In physics, temperature is clearly defined as one of the fundamental dimensions in the International System of Units \cite{international2001international}. However, in the perceptual domain, human thermal sensitivity depends on the physical characteristics of the object in contact. Specifically, the perceived temperature depends on the thermal transfer between the object itself and the human skin. This thermal interaction involves complex mechanisms that include the thermal conductivity and heat capacity of the materials involved, as well as the surface geometry and the nature of the contact \cite{ho2018material}.

\SecondRebuttal{When modeling physical haptic properties, temperature plays a unique role on the interconnection of the haptic properties, but its impact varies for each material category \cite{grimvall1999thermophysical}. For instance, certain types of plastic or organic materials undergo significant changes in haptic properties with changes in temperature. In contrast, materials such as metal, rock, wood, ceramics and fabric remain largely unaffected within temperature ranges that humans can tolerate.}

When two objects of different materials are touched, the perceived temperature is influenced by the rate of heat transfer. For instance, metals typically feel colder than wood at the same temperature because metals have higher thermal conductivity, allowing heat to be transferred away from the skin more rapidly. This concept, known as thermal diffusivity \cite{salazar2003thermal}, explains why certain materials feel warmer or cooler upon contact despite being at the same physical temperature. Moreover, the perception of temperature is not only a function of the material properties but also involves cognitive processes. Our brain integrates these thermal cues with other sensory information to form a coherent perception of the material, influencing our ability to identify and discriminate between different objects based on their thermal properties. \Rebuttal{In \cite{chen2009exploring,tiest2010tactual,okamoto2012psychophysical,cai2020thermairglove}, a deeper look into the relationship between human haptic perception and the temperature can be found.}


\SecondRebuttal{Given the uniqueness of the temperature} with respect to other haptic properties \Rebuttal{in the context of surface modeling}, we do not delve into the details in the scope of this survey.
For further analysis \Rebuttal{about the temperature related to the whole haptic pipeline}, please refer to the following comprehensive review \cite{ho2018material}.





\section{Haptic Surface Modeling}
\label{sec:haptic_surface_modeling}

\begin{table*}[!h]
\centering
\resizebox{\textwidth}{!}{%
\begin{tabular}{@{}cccccccccc@{}}
\toprule
 &  &  &  & \multicolumn{4}{c}{\textbf{Physical features}} &  &  \\ \cmidrule(lr){5-8}
\multirow{-2}{*}{\textbf{\Rebuttal{Approach}}} & \multirow{-2}{*}{\textbf{Year}} & \multirow{-2}{*}{\textbf{Method}} & \multirow{-2}{*}{\textbf{Collected signal}} & \textbf{Bumpiness} & \textbf{Friction} & \textbf{Roughness} & \textbf{Stiffness} & \multirow{-2}{*}{\textbf{Exploration}} & \multirow{-2}{*}{\textbf{Neural-based}} \\ \midrule
 & \cellcolor[HTML]{EFEFEF}2008 & \cellcolor[HTML]{EFEFEF}\cite{hover2008data} & \cellcolor[HTML]{EFEFEF}F, P, SF ,V & \cellcolor[HTML]{EFEFEF}\xmark & \cellcolor[HTML]{EFEFEF}\cmark & \cellcolor[HTML]{EFEFEF}\xmark & \cellcolor[HTML]{EFEFEF}\cmark & \cellcolor[HTML]{EFEFEF}Human & \cellcolor[HTML]{EFEFEF}\xmark \\
 & 2009 & \cite{hover2009data} & F, P, SF ,V & \xmark & \cmark & \xmark & \cmark & Human & \xmark \\
 & \cellcolor[HTML]{EFEFEF} & \cellcolor[HTML]{EFEFEF}\cite{guruswamy2009modelling} & \cellcolor[HTML]{EFEFEF}A, F & \cellcolor[HTML]{EFEFEF}\cmark & \cellcolor[HTML]{EFEFEF}\xmark & \cellcolor[HTML]{EFEFEF}\cmark & \cellcolor[HTML]{EFEFEF}\xmark & \cellcolor[HTML]{EFEFEF}Human & \cellcolor[HTML]{EFEFEF}\cmark \\
 & 2010 & \cite{hover2010user} & F, P, V & \xmark & \xmark & \xmark & \cmark & Human & \xmark \\
 & \cellcolor[HTML]{EFEFEF} & \cellcolor[HTML]{EFEFEF}\cite{landin2010dimensional} & \cellcolor[HTML]{EFEFEF}A & \cellcolor[HTML]{EFEFEF}\cmark & \cellcolor[HTML]{EFEFEF}\cmark & \cellcolor[HTML]{EFEFEF}\cmark & \cellcolor[HTML]{EFEFEF}\cmark & \cellcolor[HTML]{EFEFEF}Human & \cellcolor[HTML]{EFEFEF}\xmark \\
 &  & \cite{lang2010measurement} & A, F & \cmark & \xmark & \cmark & \cmark & Human & \xmark \\
 & \cellcolor[HTML]{EFEFEF}2011 & \cellcolor[HTML]{EFEFEF}\cite{romano2011creating} & \cellcolor[HTML]{EFEFEF}A, F, P & \cellcolor[HTML]{EFEFEF}\cmark & \cellcolor[HTML]{EFEFEF}\cmark & \cellcolor[HTML]{EFEFEF}\cmark & \cellcolor[HTML]{EFEFEF}\cmark & \cellcolor[HTML]{EFEFEF}Human & \cellcolor[HTML]{EFEFEF}\xmark \\
 & 2012 & \cite{culbertson2012refined} & A & \cmark & \cmark & \cmark & \cmark & Human & \xmark \\
 & \cellcolor[HTML]{EFEFEF} & \cellcolor[HTML]{EFEFEF}\cite{yim2012shape} & \cellcolor[HTML]{EFEFEF}F, P & \cellcolor[HTML]{EFEFEF}\xmark & \cellcolor[HTML]{EFEFEF}\xmark & \cellcolor[HTML]{EFEFEF}\xmark & \cellcolor[HTML]{EFEFEF}\cmark & \cellcolor[HTML]{EFEFEF}Human & \cellcolor[HTML]{EFEFEF}\xmark \\
 & 2013 & \cite{culbertson2013generating} & A, F, V & \cmark & \cmark & \cmark & \cmark & Human & \xmark \\
 & \cellcolor[HTML]{EFEFEF} & \cellcolor[HTML]{EFEFEF}\cite{sianov2013data} & \cellcolor[HTML]{EFEFEF}\Rebuttal{F, P} & \cellcolor[HTML]{EFEFEF}\xmark & \cellcolor[HTML]{EFEFEF}\cmark & \cellcolor[HTML]{EFEFEF}\xmark & \cellcolor[HTML]{EFEFEF}\cmark & \cellcolor[HTML]{EFEFEF}\Rebuttal{Human} & \cellcolor[HTML]{EFEFEF}\xmark \\
 & 2014 & \cite{culbertson2014modeling} & A, F, SV & \cmark & \cmark & \cmark & \cmark & Human & \xmark \\
 & \cellcolor[HTML]{EFEFEF}2015 & \cellcolor[HTML]{EFEFEF}\cite{shin2015data} & \cellcolor[HTML]{EFEFEF}A, F, V & \cellcolor[HTML]{EFEFEF}\cmark & \cellcolor[HTML]{EFEFEF}\cmark & \cellcolor[HTML]{EFEFEF}\cmark & \cellcolor[HTML]{EFEFEF}\cmark & \cellcolor[HTML]{EFEFEF}Machine & \cellcolor[HTML]{EFEFEF}\cmark \\
 & 2016 & \cite{abdulali2016data} & A, F, P & \cmark & \cmark & \cmark & \cmark & Human & \cmark \\
 & \cellcolor[HTML]{EFEFEF} & \cellcolor[HTML]{EFEFEF}\cite{yim2016data} & \cellcolor[HTML]{EFEFEF}F, P, SV & \cellcolor[HTML]{EFEFEF}\xmark & \cellcolor[HTML]{EFEFEF}\cmark & \cellcolor[HTML]{EFEFEF}\xmark & \cellcolor[HTML]{EFEFEF}\cmark & \cellcolor[HTML]{EFEFEF}Human & \cellcolor[HTML]{EFEFEF}\cmark \\
 &  & \cite{culbertson2016importance} & A, F, V & \xmark & \cmark & \cmark & \cmark & Human & \xmark \\
 & \cellcolor[HTML]{EFEFEF}2018 & \cellcolor[HTML]{EFEFEF}\cite{abdulali2018data} & \cellcolor[HTML]{EFEFEF}A, F, P & \cellcolor[HTML]{EFEFEF}\cmark & \cellcolor[HTML]{EFEFEF}\cmark & \cellcolor[HTML]{EFEFEF}\cmark & \cellcolor[HTML]{EFEFEF}\cmark & \cellcolor[HTML]{EFEFEF}Human & \cellcolor[HTML]{EFEFEF}\cmark \\
 & 2020 & \cite{abdulali2020visually} & A, F, P & \cmark & \cmark & \cmark & \cmark & Human & \xmark \\
 & \cellcolor[HTML]{EFEFEF}2022 & \cellcolor[HTML]{EFEFEF}\cite{joolee2022deep} & \cellcolor[HTML]{EFEFEF}A, F, P & \cellcolor[HTML]{EFEFEF}\cmark & \cellcolor[HTML]{EFEFEF}\cmark & \cellcolor[HTML]{EFEFEF}\cmark & \cellcolor[HTML]{EFEFEF}\cmark & \cellcolor[HTML]{EFEFEF}Human & \cellcolor[HTML]{EFEFEF}\cmark \\
 &  & \cite{cha2022data} & F, P & \xmark & \xmark & \xmark & \cmark & Machine & \cmark \\
\multirow{-21}{*}{\textbf{\rotatebox{90}{Human-Computer Haptic}}} & \cellcolor[HTML]{EFEFEF}2023 & \cellcolor[HTML]{EFEFEF}\cite{kumar2023catboost} & \cellcolor[HTML]{EFEFEF}F, P, V & \cellcolor[HTML]{EFEFEF}\xmark & \cellcolor[HTML]{EFEFEF}\xmark & \cellcolor[HTML]{EFEFEF}\xmark & \cellcolor[HTML]{EFEFEF}\cmark & \cellcolor[HTML]{EFEFEF}Machine & \cellcolor[HTML]{EFEFEF}\cmark \\ \midrule
 & 2016 & \cite{yuan2016estimating} & Img & \xmark & \xmark & \xmark & \cmark & Human/Machine & \xmark \\
 & \cellcolor[HTML]{EFEFEF}2017 & \cellcolor[HTML]{EFEFEF}\cite{yuan2017gelsight} & \cellcolor[HTML]{EFEFEF}Img & \cellcolor[HTML]{EFEFEF}\cmark & \cellcolor[HTML]{EFEFEF}\cmark & \cellcolor[HTML]{EFEFEF}\cmark & \cellcolor[HTML]{EFEFEF}\xmark & \cellcolor[HTML]{EFEFEF}Human & \cellcolor[HTML]{EFEFEF}\xmark \\
 &  & \cite{yuan2017shape} & Vid & \xmark & \xmark & \xmark & \cmark & Human/Machine & \cmark \\
 & \cellcolor[HTML]{EFEFEF}2021 & \cellcolor[HTML]{EFEFEF}\cite{cui2021self} & \cellcolor[HTML]{EFEFEF}Img & \cellcolor[HTML]{EFEFEF}\cmark & \cellcolor[HTML]{EFEFEF}\xmark & \cellcolor[HTML]{EFEFEF}\cmark & \cellcolor[HTML]{EFEFEF}\xmark & \cellcolor[HTML]{EFEFEF}Human & \cellcolor[HTML]{EFEFEF}\cmark \\
 &  & \cite{du20223d} & Img & \cmark & \xmark & \cmark & \xmark & Machine & \cmark \\
 & \cellcolor[HTML]{EFEFEF} & \cellcolor[HTML]{EFEFEF}\cite{do2022densetact} & \cellcolor[HTML]{EFEFEF}Img & \cellcolor[HTML]{EFEFEF}\cmark & \cellcolor[HTML]{EFEFEF}\xmark & \cellcolor[HTML]{EFEFEF}\cmark & \cellcolor[HTML]{EFEFEF}\xmark & \cellcolor[HTML]{EFEFEF}Machine & \cellcolor[HTML]{EFEFEF}\cmark \\
\multirow{-7}{*}{\textbf{\rotatebox{90}{Machine Haptic}}} & 2023 & \cite{do2023densetact} & Img & \cmark & \xmark & \cmark & \xmark & Machine & \cmark \\ \midrule
 & \cellcolor[HTML]{EFEFEF}2018 & \cellcolor[HTML]{EFEFEF}\cite{shin2018geometry} & \cellcolor[HTML]{EFEFEF}Img, F, A & \cellcolor[HTML]{EFEFEF}\cmark & \cellcolor[HTML]{EFEFEF}\cmark & \cellcolor[HTML]{EFEFEF}\cmark & \cellcolor[HTML]{EFEFEF}\cmark & \cellcolor[HTML]{EFEFEF}Machine & \cellcolor[HTML]{EFEFEF}\xmark \\
 & 2020 & \cite{shin2020hybrid} & Img, F, A & \cmark & \cmark & \cmark & \cmark & Machine & \xmark \\
\multirow{-3}{*}{\textbf{\rotatebox{90}{\shortstack{Mixed \\ Haptic}}}} & \cellcolor[HTML]{EFEFEF} & \cellcolor[HTML]{EFEFEF}\cite{heravi2020learning} & \cellcolor[HTML]{EFEFEF}Img, V, F & \cellcolor[HTML]{EFEFEF}\cmark & \cellcolor[HTML]{EFEFEF}\cmark & \cellcolor[HTML]{EFEFEF}\cmark & \cellcolor[HTML]{EFEFEF}\cmark & \cellcolor[HTML]{EFEFEF}Not specified & \cellcolor[HTML]{EFEFEF}\cmark \\ \bottomrule
\end{tabular}%
}
\caption{\Rebuttal{We provide a classification of data-driven approaches to haptic surface modeling, highlighting the collected signals, the physical features measured, the exploration method and the neural or deterministic nature of the algorithm used to map the collected signals to the target physical features.} A = Acceleration, F = Force, P = Position, SF = Sliding Force, V = Velocity, SV = Sliding Velocity, \Rebuttal{Img = Image, Vid = Video}}
\label{tab:haptic_modeling}
\end{table*}

\Rebuttal{Haptic surface modeling consists of mapping the physical properties that characterize a surface onto the corresponding physical feature space of choice (i.e., roughness, bumpiness, stiffness and friction). In this context, properties like force and shape are not directly relevant, as they serve to model perceptual features.} Along the modeling, physical features may require some processing, filtering, and adaptation to be transmitted to an haptic device and eventually rendered in the perception space. 

Being one of the main building blocks between the physical space and the perception, haptic surface modeling plays a crucial role in delivering a satisfactory experience. 
In this section we delve into the different methods proposed in recent research to model the haptic characteristics of a surface. 

\begin{figure}[!h]
    \centering
    \includegraphics[width=0.95\linewidth]{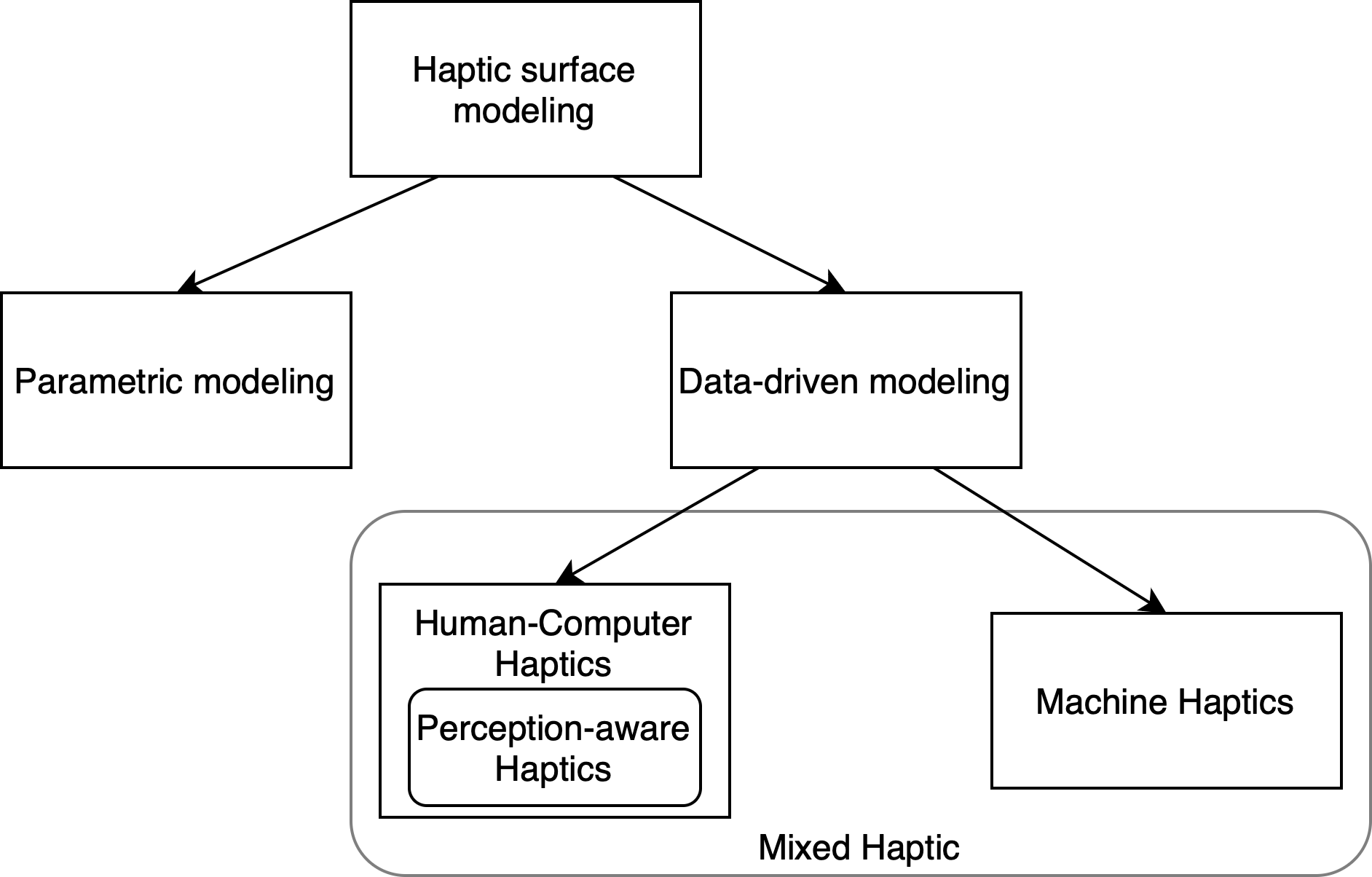}
    \caption{Haptic surface modeling approaches can be categorized into parametric and data-driven methodologies. Furthermore, data-driven methods can be subdivided into human-computer haptics, which model haptic properties with a focus on human-based applications, and machine haptics, which model haptic properties specifically for robotic applications. }
    \label{fig:haptic_modeling}
\end{figure}

As shown in Fig. \ref{fig:haptic_modeling}, existing modeling approaches can be divided in two main categories: parametric approaches \cite{basdogan2002haptic}, where the object surface is modeled as a parametric function and the more recent data-driven approaches \cite{costes2020towards}, where the mapping between a surface and the relevant physical features is learnt from real data samples.

\subsection{Parametric approaches}
Parametric approaches involve using physical, mathematical, and mechanical principles to simulate the surfaces properties. Due to their simplicity, explicit models were popular in the early studies about haptic rendering \cite{basdogan2002haptic,pacchierotti2015cutaneous,okamura2001reality}\Rebuttal{\cite{saga2013simultaneous}}. The parameters of the modeling functions can be set through measurements taken from real objects. A glance into these methods can be found in \cite{basdogan2002haptic}. Notwithstanding their attractive features, parametric models show two main limitations. Firstly, they require some a-priori knowledge of the physical phenomenon to be simulated, and this is not always available. Secondly, except for some simple cases, they require considerable sophistication to make them suitable for real-time rendering, where complex and realistic scenes with non-linear behaviors lead to heavy computation \cite{peterlik2007algorithm}. Due to these problems, such methods have found limited application up to now, and researchers have moved towards data-driven approaches, which do not require any a-priori knowledge of the underlying physical world.

\subsection{Data-driven approaches}
The general pipeline of data-driven approaches is to collect the signal generated by a sensor that explores the surface, and map it into a physical counterpart in the feature space.
The mapping function is directly learnt from real data samples belonging to a given category. Notably, the learned mapping should be successfully applied to model the haptic features of unseen objects, thus demonstrating its generalization ability.

\Rebuttal{We adopt a data-driven approach to modeling, which needs specifying how the data are collected — or rather, explored. The method of exploration affects the meaning and characteristics of the collected data. On the one hand, human exploration involves interacting with the object by applying forces and manipulating it based on pre-existing knowledge of its material and consistency. This type of exploration provides insight into how a final user might perceive and interact with the object.
On the other hand, machine exploration applies consistent forces and exploration modalities, irrespective of the object being explored. This results in a more objective and standardized measure of the object’s properties.
Thus, human exploration is more suited to capturing subjective perceptions, while machine exploration offers a uniform, objective assessment of the object.}

\Rebuttal{Another key aspect of data-driven approaches is the algorithm that allow to model the collected signals (e.g. force, position, acceleration) into the physical features of a given surface. These algorithms can be either deterministic or neural-based. Deterministic algorithms usually rely on combination of filters like the Auto-Regressive Moving Average (ARMA) to combine multiple signals into a polynomial to model physical features. Neural-based algorithms have employed classic deep network designs like auto-encoders, Convolutional Neural Networks (CNN) to map input (mostly vision-based like images and videos) to physical features. While deterministic approaches provide more control over the process and enhanced explainability, neural-base approaches have shown superior abilities in capturing the relations between the collected signals and physical features.}

Data-driven approaches, reported in Tab. \ref{tab:haptic_modeling}, can be classified as either \textbf{computer haptic}, which refers to human-computer interaction applications, \textbf{machine haptic}, which refers to robot-based applications, or \textbf{mixed haptic}, which mixes data usually employed by computer haptic or machine haptic, depending on the collected signal.

\begin{figure}[!h]
\centering
\subfigure[]{\includegraphics[width=0.95\linewidth]{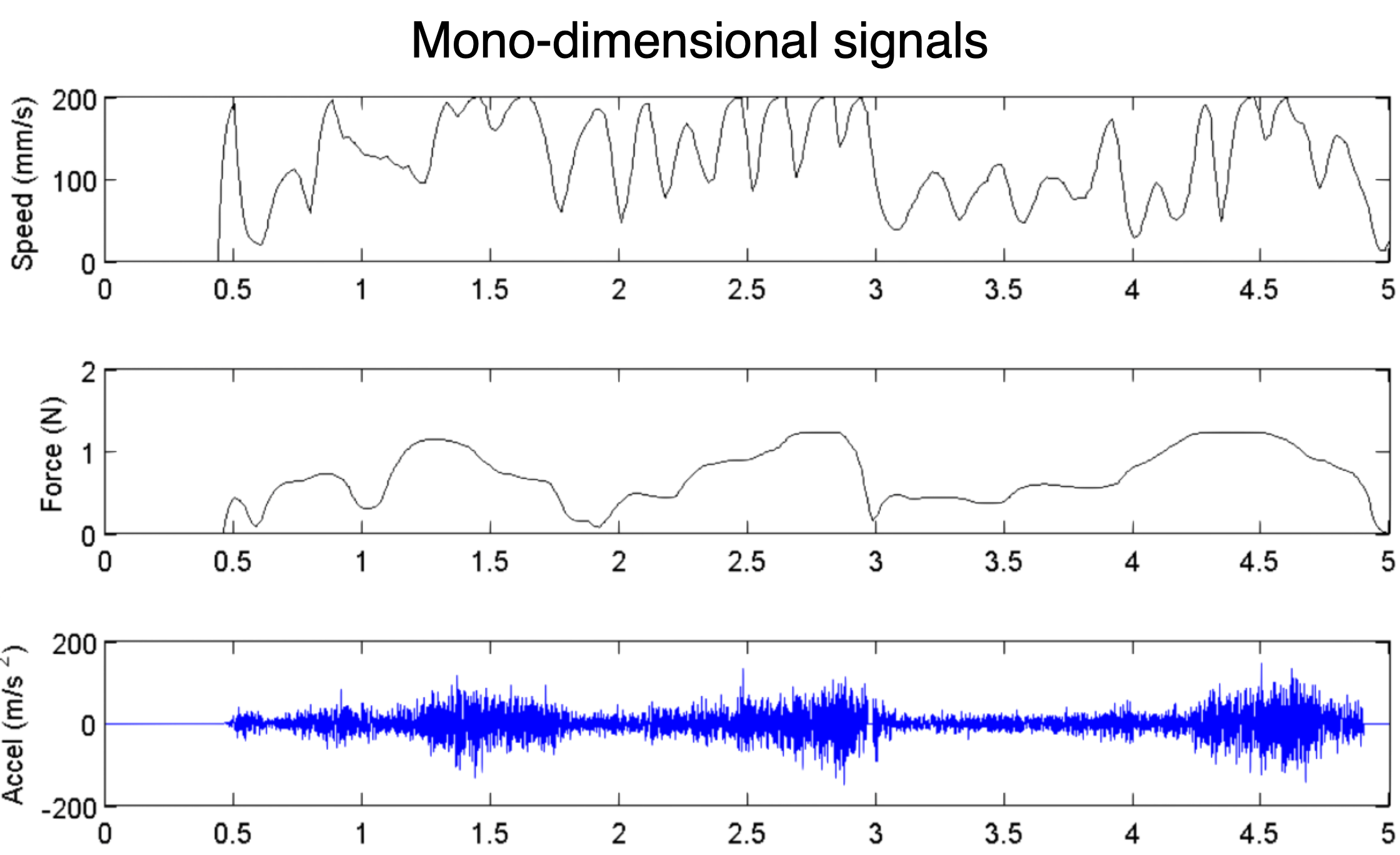}}
\subfigure[]{\includegraphics[width=0.95\linewidth]{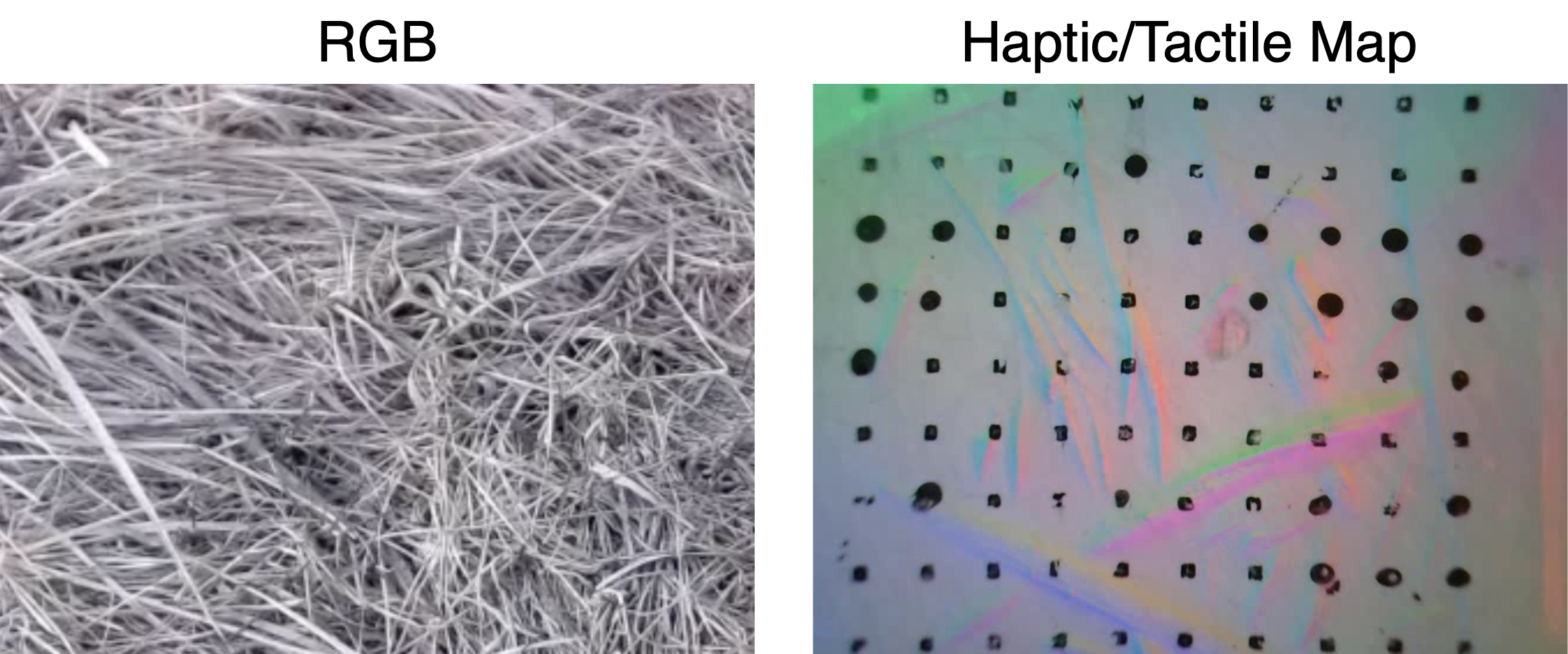}}
\caption{Data-driven approaches require the collection of haptic signals, which can be either (a) mono-dimensional (taken from \cite{culbertson2012refined}) or (b) vision-based (taken from \cite{yang2022touch}). Mono-dimensional signals usually describe a single feature acquired by a sensor, like speed, force, and acceleration. Vision-based signals provide spatially-related 2D high-resolution information of a surface.}
\label{fig:data_collected}
\end{figure}

\paragraph{Human-computer haptic signals} 
Applications based on the so-called computer haptic aim at modeling the haptic properties of a surface by mimicking the interaction between the surface and the human finger \cite{hover2009data}. The usual pipeline involves the use of sensors, which touch the surface of interest and produce one or multiple one-dimensional signals corresponding to the haptic information, as shown in Fig. \ref{fig:data_collected} (a). The most widely used sensors are the inertial ones, such as accelerometers, probes and force sensors.

The collected signals depend on the motion and position of the sensor when touching the object: \textit{position}, meaning the displacement of the sensing device, \textit{velocity}, namely how fast the sensing device moves, \textit{sliding velocity}, which refer to how fast the sensing device moves on the plane tangential to the surface and, similarly, \textit{acceleration}, \textit{force}, and \textit{sliding force} to which the device is subject. 

As shown in the upper part of Tab. \ref{tab:haptic_modeling}, various studies have explored different one-dimensional signals translating them into the physical features. In \cite{hover2008data}, the authors focused on modeling visco-elastic properties of surfaces, particularly silicon-based and rubber materials. They measured position and force applied simultaneously, using interpolation methods to model transient behaviors. In \cite{hover2009data}, they utilized a Generalized Maxwell Model for elasticity, focusing on visco-elastic bodies and viscous fluids. In \cite{guruswamy2009modelling}, they examined bumpiness and roughness using acceleration and force data collected with a WhAT probe, applying an IIR filter for their modeling. The work in \cite{hover2010user} assessed human force sensing capacity and evaluated their previous modeling method \cite{hover2008data} by testing on two silicon cylinders. The authors used both static and dynamic models based on radial basis function interpolation. In \cite{landin2010dimensional}, the authors reduced a 3D high-frequency acceleration signal to a 1D signal using their DFT321 transformation. This method approximates the 3D signal by aligning the module and phase through Discrete Time Fourier Transformation. In \cite{lang2010measurement}, they used a sensor fusion technique combining haptic data from a probe with visual data from a camera. They applied the Hertzian contact model to describe the stiffness of objects, achieving a 3D profile from the collected acceleration and force data. Building on \cite{landin2010dimensional}, \cite{romano2011creating} used Linear Predictive Coding (LPC) to model time-domain patterns from 3D data reduced to 1D, creating a look-up table for data prediction. Subsequently, in \cite{culbertson2012refined}, they enhanced the method from \cite{romano2011creating} using an Auto-Regressive Moving Average (ARMA) model, combining past outputs and weighted moving averages of past inputs to associate time-domain patterns with spatial domains. \Rebuttal{The study by \cite{yim2012shape} relied on a PHANToM Premium 1.5 High Force device and an ATI Nano 17 force/torque sensor to model the shape variation (stiffness) of soft objects by leveraging force and position collected and using a sum of square errors (SSE) fitting function. A second approach employing a similar hardware, presented by \cite{sianov2013data}, measures force and position to model friction and stiffness.} 
In \cite{culbertson2013generating}, data were collected with users moving freely along surfaces. The authors used an Auto-Regressive model, assuming signal stationarity and segmenting signals into chunks using an Auto-PARM algorithm. In \cite{culbertson2014modeling}, they continued with DFT321, using an AR model and Auto-PARM algorithm for signal segmentation, and employed Delaunay triangulation for interpolation. The authors of \cite{shin2015data} modeled isotropic haptic textures with a frequency-decomposed neural network, demonstrating its applicability to anisotropic textures. The data were collected using a Texanner and processed with a neural network model. In \cite{abdulali2016data}, they segmented the data using a bottom-up algorithm and interpolated it with a Radial Basis Function Network (RBFN) to estimate Line Spectrum Frequency coefficients. They analyzed homogeneous anisotropic grain textures, scanning in two directions. In \cite{yim2016data}, the authors used a force sensor and PHANToM device, developing an interpolation model based on the proxy position of an imaginary contact point. They identified the sliding yield surface through automated palpation. The authors in \cite{culbertson2016importance} incorporated Dahl friction model for friction, labeled stiffness with the speed of the tooltip, and used an AR model for roughness. They rendered virtual surfaces with a haptic interface augmented with a Tactile Labs Haptuator. In \cite{abdulali2018data} they followed the segmentation and interpolation methods from their previous work, analyzing bumpiness, friction, roughness, and stiffness using an accelerometer, force sensor, and PHANToM device. Next, the same authors, in \cite{abdulali2020visually}, proposed another method to collect and process data. They demonstrated that by combining their segmentation framework and a user interface, which guides the data collection process to create a human-in-loop system, the approximation quality of the model increases. In \cite{joolee2022deep}, they trained a neural network with contact acceleration data collected through a manual scanning stylus. They used attention-aware 1D CNNs and encoder-decoder networks with Bi-LSTM to capture spatial and temporal dynamics. The authors from \cite{cha2022data} used a force-feedback device and load cell to measure force and position, employing Radial Basis Function and Random Forest models for interpolation.
Lastly, in \cite{kumar2023catboost} they utilized a CatBoost approach, which is a variant of gradient boosting. Here they train decision trees to learn the required mapping function for modeling the objects.
A notable mention should be made of the work presented in \cite{landin2010dimensional, romano2011creating}, which introduced the DFT321 method. This technique substantially contributed to the modeling of haptic properties and led to the creation of the first dataset containing data on surface haptic properties \cite{culbertson2014one}. However, the DFT321 approach relies on lookup tables to model surfaces, which limits its ability to generalize effectively to new, unseen surfaces. In contrast, machine learning techniques offer enhanced generalization capabilities, enabling more robust and versatile modeling of haptic properties across a broader range of surfaces.

Concluding, human-computer haptic signals have long been the focus of Haptic Surface Modeling thanks to their simple processing techniques, which allows to immediately characterize the features of a specific point on the surface. However, the same sensed signals (e.g., acceleration, force) have been modeled into different physical features depending on the design of each sensor and the dataset being used. This makes it difficult for researchers to perform comparative studies. As an example, relating how different sensors characterize the physical features of the same surface would help in comparing the different modeling techniques. The other main limitation of human-computer approaches is the difficult in geometrically relating different points in space.

\Rebuttal{\paragraph{Perception-aware haptic signals} A recently proposed approach in human-computer haptic modeling involves incorporating the user into the modeling process. Unlike previous methods, this set of techniques focuses on modeling haptic properties by taking users' perceptions into account, relying directly on human judgment to guide the process. One of the earliest works using this approach is presented in \cite{piovarvci2016interaction}. The authors aim to define a model that maps physical stiffness, characterized by surface displacement under applied pressure, to the stiffness perceived by users. To achieve this, they first established a perceptual space that groups similar stiffness-related stimuli. Then, they employed a non-linear model, originally introduced in \cite{pressman2006perception}, augmented with terms that account for pressing force to identify the relation between the physical stiffness and its perceived counterpart. In a subsequent study presented in \cite{piovarvci2018perception}, the authors constructed a perceptual space for friction and applied a non-metric Multi-Dimensional Scaling (MDS) algorithm, as defined in \cite{wills2009toward}, to establish a correlation between the physical property and the perceptual space. Differently from previous method, the authors in \cite{tymms2018quantitative} defined a perceptual space for roughness and bumpiness and create a data-driven model, based on height maps, to map the physical stimuli to the perceptual dimensions. In \cite{degraen2021capturing}, the authors sampled 15 different tissues using GelSight sensors and synthetically replicated their characteristics through 3D printing. The study aimed to validate the modeling process by assessing the perceived similarity between the real tissues and their synthetic counterparts. Similarly, in \cite{kurita2023generation}, the authors developed a data-driven model for roughness and bumpiness, directly guided by user input. This approach entailed adjusting the model's input according to user evaluations. In contrast to the previous approaches, other methods aim to model the haptic physical properties indirectly. In \cite{verschoor2020tactile}, the authors propose a simulation framework where an hand skin deformation model, guided from a data-driven method, is employed to identify the haptic stimuli and render them to the user. In \cite{shao2023haptic} proposed an indirect method for modeling the haptic properties of surfaces by designing a cognitive model that characterizes how humans interact with physical surfaces.}

\Rebuttal{Involving the user in the loop to model haptic properties is a promising approach for understanding how haptic stimuli can be replicated. However, these methods are challenging to classify and compare with previous approaches, as they do not rely on specific signals to model the properties. Instead, they operate within a psychophysical dimension, requiring distinct evaluation methods that fall outside the scope of this work.}

\paragraph{Machine haptic signals} While human-computer haptic applications have been developed to work with humans using one-dimensional signals, machine haptics aims at making a robotic hand deal with objects to accomplish different tasks, such as object manipulation \cite{luo2017robotic}. To achieve this goal, object geometry must be made available. Therefore, machine haptic generally relies on vision-based sensors to obtain 2D high-resolution information in the form of RGB images and haptic/tactile maps, as shown in Fig. \ref{fig:data_collected} (b). \Rebuttal{While these maps can be seen as representations of monodimensional signals, they represent spatially related 1D signals, offering a multidimensional perspective for exploring the collected haptic data.} The sensors capturing these data rely on a RGB camera recording the deformation of a membrane upon its contact with a surface. The images of the deformed membrane can be directly used as an haptic map or they can be used to estimate the depth of an object representing its shape. 
In \cite{yuan2016estimating}, a method for estimating object hardness using a touch sensor and linear regression is proposed, introducing the GelSight sensor. The development and application of the GelSight sensor are thoroughly reviewed in \cite{yuan2017gelsight}. This paper covers the sensor's optical system design and algorithms for deriving the haptic physical features from the images. GelSight’s performance in geometry and force measurement is experimentally evaluated, demonstrating its capabilities in capturing friction, roughness, and bumpiness. In \cite{yuan2017shape}, a neural network is employed to map an image sequence to a scalar hardness value. This approach measures the hardness of objects with different shapes under loosely controlled contact conditions, utilizing both manual and robotic interactions. In \cite{cui2021self} the authors introduce the GelStereo sensor and the method that converts tactile information into geometric deformation of a silicon gel layer, captured using visual sensing methods. The study employs a self-supervised GS-DepthNet estimation technique to enhance the accuracy and resolution of the contact geometry, focusing on roughness and bumpiness. The DelTact sensor, discussed in \cite{du20223d}, uses a Gaussian density regression model to estimate roughness and bumpiness from an image. The model solves a point cloud that aligns with prior depth information, ensuring depth consistency and generalizing across sensors with minimal parameter tuning. In \cite{do2022densetact}, the DenseTact model estimates depth images from RGB input images using a DenseNet. This method aims to capture roughness and bumpiness by converting tactile information into depth images. The improved DenseTact2.0 sensor is presented in \cite{do2023densetact}, which includes an encoder-decoder network for shape estimation and force estimation models using transfer learning.

Differently from human-computer haptic signals, machine haptic applications leverage the larger expressiveness of images to represent the data acquired, introducing a new aspect to haptic sensing. By directly sensing the surface geometry, it is possible to achieve higher precision and richer data representation. This capability allows robotic systems to better understand and interact with their environments, facilitating more complex tasks such as detailed object manipulation, texture recognition, and adaptive control based on the tactile feedback received.

\paragraph{Mixed haptic modeling methods} Human-computer and machine haptic applications have traditionally been studied separately due to their different requirements, each facing limitations such as difficult in scaling to the large variety of real materials or inadequate modeling of skin-material interactions. Recently, researchers have proposed methods that integrate the two approaches to model the haptic properties of surfaces. These methods use vision-based sensors or standard cameras to estimate 2D geometric data and inertial sensors for stiffness and friction.
In \cite{shin2018geometry}, the authors combined data-driven and parametric modeling to estimate the haptic properties of surfaces. They used a photometric stereo algorithm to estimate the bidirectional reflectance distribution function (BRDF), modeling surface texture through data-driven methods for roughness and bumpiness, and physical-based approaches for stiffness and friction. In \cite{shin2020hybrid}, the model is enhanced by incorporating an LPC-based method to capture vibrations. By merging force and vibration feedback models based on their spectral characteristics, a comprehensive hybrid framework is devised, which allows for more accurate 3D surface reconstruction. In \cite{heravi2020learning}, the authors enhanced the data collected in \cite{culbertson2014one} by employing a GelSight sensor to refine the samples of some materials with additional geometric information. Their architecture encodes the GelSight images into texture representation vectors, which are then combined with encoded action representations derived from the user’s force and velocity. This combined representation is used by an acceleration predictor module to predict the desired DFT (Discrete Fourier Transform).

\section{Datasets and Simulators}
\label{sec:datasets}

\begin{table*}[]
\centering
\resizebox{\textwidth}{!}{%
\begin{tabular}{cccccccccccccccc}
\hline
                                &                                 &                                 & \multicolumn{5}{c}{\textbf{Haptic data}}                                                                                                                                                                                                                                                                             &                          & \multicolumn{3}{c}{\Rebuttal{\textbf{Other data}}}                                                                                                                                & \textbf{}           & \multicolumn{2}{c}{\textbf{Dataset size}} & \textbf{}                                                                                 \\ \cline{4-8} \cline{10-12} \cline{14-15}
\multirow{-2}{*}{\textbf{Year}} & \multirow{-2}{*}{\textbf{Name}} & \multirow{-2}{*}{\textbf{Ref.}} & \textbf{\begin{tabular}[c]{@{}c@{}}Haptic \\ map\end{tabular}} & \textbf{\begin{tabular}[c]{@{}c@{}}Acceleration/\\ vibration\end{tabular}} & \textbf{Position} & \textbf{\begin{tabular}[c]{@{}c@{}}Pressure\\ Force\end{tabular}} & \textbf{\begin{tabular}[c]{@{}c@{}}Friction force/\\ coefficient\end{tabular}} & \textbf{}                & \textbf{\begin{tabular}[c]{@{}c@{}}RGB\\ image\end{tabular}} & \textbf{Sound}                      & \textbf{\begin{tabular}[c]{@{}c@{}}Depth/\\ 3D data\end{tabular}} & \textbf{Real/Synth} & \textbf{Objects}    & \textbf{Touches}    & \textbf{Link}                                                                             \\ \hline
\rowcolor[HTML]{EFEFEF} 
2014                            & HaTT                            & \cite{culbertson2014one}        &                                                                & \cmarkblack                                                                & \cmarkblack       & \cmarkblack                                                       &                                                                                & \cellcolor[HTML]{EFEFEF} & \cellcolor[HTML]{EFEFEF}\cmarkblack                          & \cellcolor[HTML]{EFEFEF}            & \cellcolor[HTML]{EFEFEF}                                          & Real                & 100                 & -                   & \href{http://repository.upenn.edu/meam_papers/299/}{Link}                                 \\
                                & LMT v1                          & \cite{strese2014haptic}         &                                                                & \cmarkblack                                                                &                   &                                                                   &                                                                                &                          &                                                              &                                     &                                                                   & Real                & 43                  & -                   & \href{https://zeus.lmt.ei.tum.de/downloads/texture/}{Link}                                \\
\rowcolor[HTML]{EFEFEF} 
2015                            & {\color[HTML]{333333} LMT v2}   & \cite{strese2015surface}        &                                                                & \cmarkblack                                                                &                   &                                                                   &                                                                                & \cellcolor[HTML]{EFEFEF} & \cellcolor[HTML]{EFEFEF}                                     & \cellcolor[HTML]{EFEFEF}\cmarkblack & \cellcolor[HTML]{EFEFEF}                                          & Real                & 69                  & -                   & \href{https://zeus.lmt.ei.tum.de/downloads/texture/}{Link}                                \\
2016                            & LMT v3                          & \cite{strese2016multimodal}     &                                                                & \cmarkblack                                                                &                   &                                                                   & \cmarkblack                                                                    &                          & \cmarkblack                                                  & \cmarkblack                         &                                                                   & Real                & 69                  & -                   & \href{https://zeus.lmt.ei.tum.de/downloads/texture/}{Link}                                \\
\rowcolor[HTML]{EFEFEF} 
2017                            & FoS                             & \cite{calandra2017feeling}      & \cmarkblack                                                    &                                                                            &                   &                                                                   &                                                                                & \cellcolor[HTML]{EFEFEF} & \cellcolor[HTML]{EFEFEF}\cmarkblack                          & \cellcolor[HTML]{EFEFEF}            & \cellcolor[HTML]{EFEFEF}                                          & Real                & 106                 & 9.2k                & \href{https://sites.google.com/view/the-feeling-of-success/}{Link}                        \\
                                & Fabrics                         & \cite{yuan2017connecting}       & \cmarkblack                                                    &                                                                            &                   &                                                                   &                                                                                &                          & \cmarkblack                                                  &                                     & \cmarkblack                                                       & Real                & 118                 & -                   & \href{http://people.csail.mit.edu/yuan_wz/fabricdata/GelFabric.tar.gz}{Link}              \\
\rowcolor[HTML]{EFEFEF} 
                                & UHL                             & \cite{hassan2017towards}        &                                                                &                                                                            &                   & \cmarkblack                                                       & \cmarkblack                                                                    & \cellcolor[HTML]{EFEFEF} & \cellcolor[HTML]{EFEFEF}\cmarkblack                          & \cellcolor[HTML]{EFEFEF}            & \cellcolor[HTML]{EFEFEF}                                          & Real                & 84                  & -                   & \href{http://haptics.khu.ac.kr/thehapticlibrary/}{Link}                                   \\
                                & Texplorer                       & \cite{strese2017content}        &                                                                & \cmarkblack                                                                &                   &                                                                   & \cmarkblack                                                                    &                          & \cmarkblack                                                  & \cmarkblack                         &                                                                   & Real                & 108                 & -                   & \href{https://zeus.lmt.ei.tum.de/downloads/texture/}{Link}                                \\
\rowcolor[HTML]{EFEFEF} 
2018                            & ViTac                           & \cite{luo2018vitac}             & \cmarkblack                                                    &                                                                            &                   &                                                                   &                                                                                & \cellcolor[HTML]{EFEFEF} & \cellcolor[HTML]{EFEFEF}\cmarkblack                          & \cellcolor[HTML]{EFEFEF}            & \cellcolor[HTML]{EFEFEF}                                          & Real                & 100                 & 3.0k                & \href{https://drive.google.com/file/d/1uYy4JguBlEeTllF9Ch6ZRixsTprGPpVJ/view?pli=1}{Link} \\
                                & MTaF                            & \cite{calandra2018more}         & \cmarkblack                                                    &                                                                            &                   &                                                                   &                                                                                &                          & \cmarkblack                                                  &                                     &                                                                   & Real                & 65                  & 6.45k               & \href{https://drive.google.com/drive/folders/1wHEg_RR8YAQjMnt9r5biUwo5z3P6bjR3}{Link}     \\
\rowcolor[HTML]{EFEFEF} 
2019                            & Haptex                          & \cite{jiao2019haptex}           &                                                                &                                                                            & \cmarkblack       & \cmarkblack                                                       & \cmarkblack                                                                    & \cellcolor[HTML]{EFEFEF} & \cellcolor[HTML]{EFEFEF}                                     & \cellcolor[HTML]{EFEFEF}            & \cellcolor[HTML]{EFEFEF}                                          & Real                & 120                 & -                   & \href{https://haptic.buaa.edu.cn/English_FabricDatabase.htm}{Link}                        \\
                                & VisGel                          & \cite{li2019connecting}         & \cmarkblack                                                    &                                                                            &                   &                                                                   &                                                                                &                          & \cmarkblack                                                  &                                     &                                                                   & Real                & 195                 & 12.0k               & \href{http://visgel.csail.mit.edu}{Link}                                                  \\
\rowcolor[HTML]{EFEFEF} 
                                & Texplorer2                      & \cite{strese2019haptic}         &                                                                & \cmarkblack                                                                &                   &                                                                   & \cmarkblack                                                                    & \cellcolor[HTML]{EFEFEF} & \cellcolor[HTML]{EFEFEF}\cmarkblack                          & \cellcolor[HTML]{EFEFEF}\cmarkblack & \cellcolor[HTML]{EFEFEF}                                          & Real                & 184                 & -                   & \href{https://zeus.lmt.ei.tum.de/downloads/texture/}{Link}                                \\
2020                            & HaTT + Gelsight                 & \cite{heravi2020learning}       & \cmarkblack                                                    & \cmarkblack                                                                & \cmarkblack       & \cmarkblack                                                       &                                                                                &                          & \cmarkblack                                                  &                                     &                                                                   & Real                & 100                 & -                   & \href{https://sites.google.com/stanford.edu/haptic-texture-generation/dataset}{Link}      \\
\rowcolor[HTML]{EFEFEF} 
2021                            & ObjectFolder                    & \cite{gao2021objectfolder}      &                                                                &                                                                            & \cmarkblack       &                                                                   &                                                                                & \cellcolor[HTML]{EFEFEF} & \cellcolor[HTML]{EFEFEF}\cmarkblack                          & \cellcolor[HTML]{EFEFEF}\cmarkblack & \cellcolor[HTML]{EFEFEF}\cmarkblack                               & Synth               & 100                 & -                   & \href{https://objectfolder.stanford.edu}{Link}                                            \\
2022                            & Touch and Go                    & \cite{yang2022touch}            & \cmarkblack                                                    &                                                                            &                   &                                                                   &                                                                                &                          & \cmarkblack                                                  &                                     &                                                                   & Real                & 3971                & 13.9k               & \href{https://touch-and-go.github.io}{Link}                                               \\
\rowcolor[HTML]{EFEFEF} 
                                & ObjectFolder2.0                 & \cite{gao2022objectfolder}      &                                                                &                                                                            & \cmarkblack       &                                                                   &                                                                                & \cellcolor[HTML]{EFEFEF} & \cellcolor[HTML]{EFEFEF}\cmarkblack                          & \cellcolor[HTML]{EFEFEF}\cmarkblack & \cellcolor[HTML]{EFEFEF}\cmarkblack                               & Synth               & 1000                & -                   & \href{https://objectfolder.stanford.edu}{Link}                                            \\
                                & PoseIt                          & \cite{kanitkar2022poseit}       & \cmarkblack                                                    &                                                                            &                   & \cmarkblack                                                       &                                                                                &                          & \cmarkblack                                                  &                                     &                                                                   & Real                & 26                  & 1.84k               & \href{https://github.com/CMURoboTouch/PoseIt}{Link}                                       \\
\rowcolor[HTML]{EFEFEF} 
                                & SSVTP                           & \cite{kerr2022self}             & \cmarkblack                                                    &                                                                            &                   &                                                                   &                                                                                & \cellcolor[HTML]{EFEFEF} & \cellcolor[HTML]{EFEFEF}\cmarkblack                          & \cellcolor[HTML]{EFEFEF}            & \cellcolor[HTML]{EFEFEF}                                          & Real                & 10                  & 4.5k                & \href{https://drive.google.com/file/d/1H0B-jJ4l3tJu2zuqf-HbZy2bjEl-vL3f/view}{Link}       \\
2023                            & ObjectFolder Real               & \cite{gao2023objectfolder}      & \cmarkblack                                                    &                                                                            &                   &                                                                   &                                                                                &                          & \cmarkblack                                                  & \cmarkblack                         & \cmarkblack                                                       & Real                & 100                 & -                   & \href{https://objectfolder.stanford.edu}{Link}                                            \\
\rowcolor[HTML]{EFEFEF} 
                                & MTTD                            & \cite{lima2023multimodal}       &                                                                & \cmarkblack                                                                &                   &                                                                   &                                                                                & \cellcolor[HTML]{EFEFEF} & \cellcolor[HTML]{EFEFEF}                                     & \cellcolor[HTML]{EFEFEF}            & \cellcolor[HTML]{EFEFEF}                                          & Real                & 12                  & 3.6k                & \href{https://data.mendeley.com/datasets/n666tk4mw9/1}{Link}                              \\
                                & YCB-Slide                       & \cite{suresh2023midastouch}     & \cmarkblack                                                    &                                                                            &                   &                                                                   &                                                                                &                          & \cmarkblack                                                  &                                     & \cmarkblack                                                       & Real/Synth          & 10                  & -                   & \href{https://github.com/rpl-cmu/YCB-Slide?tab=readme-ov-file\#Dataset-details}{Link}     \\
\rowcolor[HTML]{EFEFEF} 
2024                            & MPI-10                          & \cite{khojasteh2024mpi10}       &                                                                & \cmarkblack                                                                &                   & \cmarkblack                                                       &                                                                                & \cellcolor[HTML]{EFEFEF} & \cellcolor[HTML]{EFEFEF}                                     & \cellcolor[HTML]{EFEFEF}\cmarkblack & \cellcolor[HTML]{EFEFEF}                                          & Real                & 10                  & -                   & \href{https://edmond.mpg.de/dataset.xhtml?persistentId=doi:10.17617/3.PM8R94}{Link}       \\
                                & SSVTP + HCT\textsuperscript{*}  & \cite{fu2024touch}              & \cmarkblack                                                    &                                                                            &                   &                                                                   &                                                                                &                          & \cmarkblack                                                  &                                     &                                                                   & Real                & -                   & 44.0k               & \href{https://huggingface.co/datasets/mlfu7/Touch-Vision-Language-Datase}{Link}           \\
\rowcolor[HTML]{EFEFEF} 
                                & SENS3\textsuperscript{T}        & \cite{balasubramanian2024sens3} &                                                                & \cmarkblack                                                                & \cmarkblack       & \cmarkblack                                                       & \cmarkblack                                                                    & \cellcolor[HTML]{EFEFEF} & \cellcolor[HTML]{EFEFEF}\cmarkblack                          & \cellcolor[HTML]{EFEFEF}\cmarkblack & \cellcolor[HTML]{EFEFEF}                                          & Real                & 50                  & -                   & \href{https://www.sens3.net/}{Link}                                                       \\
                                & Touch100k\textsuperscript{*}    & \cite{cheng2024touch100k}       & \cmarkblack                                                    &                                                                            &                   &                                                                   &                                                                                &                          & \cmarkblack                                                  &                                     &                                                                   & Real                & -                   & 100k                & \href{https://cocacola-lab.github.io/Touch100k/}{Link}                                    \\
\rowcolor[HTML]{EFEFEF} 
                                & TARF                            & \cite{dou2024tactile}           & \cmarkblack                                                    &                                                                            &                   &                                                                   &                                                                                & \cellcolor[HTML]{EFEFEF} & \cellcolor[HTML]{EFEFEF}                                     & \cellcolor[HTML]{EFEFEF}            & \cellcolor[HTML]{EFEFEF}\cmarkblack                               & Real                & -                   & 19.3k               & \href{https://github.com/Dou-Yiming/TaRF}{Link}                                           \\ \hline
\end{tabular}
}
\caption{This table provides a comprehensive overview of datasets used in haptic sensing and interaction research, highlighting their diverse properties. Each dataset is detailed with its name, key features \Rebuttal{(both strictly haptic related and other data type)}, data collection methods, and data size. The symbols (\textbf{*}) and (\textsuperscript{T}) indicate the datasets providing as data annotations text descriptions and temperature respectively.}
\label{tab:datasets}
\end{table*}

The field of haptic data acquisition and simulation has seen significant advancements, driven by the need for more realistic and immersive tactile interactions in robotics, virtual reality, and human-computer interaction. Various datasets and simulators have been developed to capture and model the complex sensory experiences associated with touch. These resources provide comprehensive data that combine visual, tactile, auditory, and sometimes linguistic information, facilitating the development and evaluation of algorithms for tasks such as texture classification, material recognition, and robotic manipulation. In this section, we summarize the main datasets available in the literature, describing their properties, data collection methods, features, and potential applications, distinguishing between synthetic and real datasets. The details are summarized in Tab. \ref{tab:datasets}. Additionally, we review recent simulators that enhance the realism of tactile data, bridging the gap between simulated and real-world haptic interactions.

\subsection{Datasets}
In \cite{culbertson2014one} (HaTT), acceleration data was collected to model vibrations from textures, position, and pressure force, along with RGB images for 100 haptic textures and friction models. The dataset, including ten-second recordings of force, speed, and high-frequency acceleration from a handheld tool, uses a lookup table where AR models are labeled by force and speed. 

The paper \cite{strese2014haptic} introduces a dataset (LMT v1) of acceleration data for 43 textures to aid in developing algorithms for texture recognition through tool-mediated interactions. It has been improved in \cite{strese2015surface} (LMT v2), introducing a novel dataset and methodology for classifying surfaces based on acceleration signals from human freehand movements and in \cite{strese2016multimodal} (LMT v3), where high-resolution images, haptic feedback, and sound recordings of various surface materials were collected. The dataset includes diverse materials like wood, metal, fabric, and plastic, with annotations for supervised learning, capturing visual features (color histograms, texture descriptors, edge detection), tactile features (vibration frequency, amplitude, force measurements), and auditory features (sound frequency, intensity, temporal patterns) to support feature fusion techniques for improved classification accuracy.

In \cite{calandra2017feeling} (FoS), a 7-DoF Sawyer robotic arm with a Weiss WSG-50 gripper and GelSight tactile sensors, along with a Microsoft Kinect 2 for depth data, was used to collect over 9000 grasping trials. Visuo-tactile deep neural networks trained on this data showed that incorporating tactile information significantly improved grasp prediction accuracy. The work in \cite{yuan2017connecting} (Fabrics) presented a dataset and methodology to bridge visual and tactile perceptions of materials using 118 fabric samples (Fabrics). Convolutional Neural Networks (CNNs) trained on this data generated embedding vectors to link visual inputs with tactile sensations, demonstrating that models incorporating both visual and tactile data outperformed those using visual data alone. The dataset from \cite{hassan2017towards} (UHL) aims to create a universal library of haptic textures assignable to surfaces based on their visual features. This library enables automatic haptic feedback based on visual appearance, enhancing realism and immersion in virtual environments, such as virtual reality and remote manipulation. \cite{strese2017content} (Texplorer) includes diverse surface materials, captured using high-resolution imaging and tactile sensors. Designed for developing and evaluating retrieval algorithms, it supports multi-modal research and is ideal for applications in digital material libraries, e-commerce, and virtual reality. In \cite{strese2019haptic} (Texplorer2), a Tactile Computer Mouse (TCM) is introduced to simulate tactile sensations like hardness, friction, warmth, and roughness, enhancing virtual material perception. Equipped with actuators, the TCM provides calibrated tactile feedback matched to visual textures across 184 different surfaces, aiming to deliver realistic material interaction experiences.

The dataset in \cite{luo2018vitac} (ViTac) comprises various cloth textures captured using both visual and tactile sensing methods. This multi-modal dataset supports research into integrating visual and tactile features for enhanced cloth texture recognition, aiding in the development and evaluation of algorithms for improved material identification and classification. In \cite{calandra2018more} (MTaF), the dataset combines high-resolution visual data capturing object appearance and shape with tactile sensor readings detailing contact forces and surface texture. 

\cite{jiao2019haptex} (Haptex) aims to enhance virtual texture realism through tactile displays, featuring data from 120 fabric samples with diverse textures. In \cite{li2019connecting} (VisGel), RGB and GelSight data from 118 surfaces are collected to explore integrating visual and tactile sensory information for enhanced physical interactions. The study focuses on training a cross-modal prediction model using conditional Generative Adversarial Networks (GANs), enabling robots to predict tactile sensations from visual inputs and vice versa.

\cite{heravi2020learning} (HaTT + GelSight) provides models for 100 materials using a sensorized pen measuring 6 degrees of freedom (DoF) in force/torque, acceleration, and positional tracking during 10-second circular motions. The dataset supports research in texture rendering and hardness estimation for enhancing haptic feedback realism in virtual reality and teleoperation applications.

The dataset presented in \cite{gao2021objectfolder} (ObjectFolder) is a dataset of multisensory neural and real objects, containing 100 virtualized objects with visual, auditory, and tactile representations. The successor of ObjectFolder is \cite{gao2022objectfolder} (ObjectFolder2.0). The purpose of the dataset is to facilitate sim-to-real transfer in computer vision and robotics. They synthesized 1000 common household objects, with implicit neural representations. The rendering time has been improved, it’s orders of magnitude faster, with improved rendering quality for visual, acoustic, and tactile modalities. Models trained on virtual objects in ObjectFolder2.0 successfully transfer to real-world counterparts. A comprehensive suite of tasks designed for multisensory object-centric learning is provided in \cite{gao2023objectfolder} (ObjectFolder Real). The benchmark focuses on object recognition, reconstruction, and manipulation using sight, sound, and touch. It emphasizes the importance of multisensory perception in computer vision and robotics. It contains real data. Measurements cover 3D meshes, videos, impact sounds, and tactile readings. Systematic benchmarking reveals the roles of vision, audio, and touch in different object-centric learning tasks. Multisensory perception significantly impacts performance.

The work by Yang et al. \cite{yang2022touch} (Touch and Go) has introduced a multimodal visuo-tactile learning dataset focusing on associating touch with sight during physical interactions with objects. Human data collectors used tactile sensors to probe objects in natural environments, simultaneously recording egocentric video. In \cite{kanitkar2022poseit} the authors demonstrated how holding poses impact grasp stability. The purpose of the dataset (PoseIt) is investigating grasp stability considering different holding poses. An LSTM classifier trained on PoseIt achieves 0.85 accuracy on this task. Multi-modal models outperform using vision or tactile data alone. Moreover, it generalizes to unseen objects and poses. 

The dataset (SSVTP) \cite{kerr2022self} enables robots to learn multi-task visuo-tactile representations in a self-supervised manner. It includes spatially-aligned visual and tactile image pairs, allowing the robot to perform tasks like feature classification, contact localization, anomaly detection, visual query-based feature search, and edge following along cloth edges.

In \cite{khojasteh2024mpi10} (MPI-10) they capture haptic-auditory recordings during human exploration of 10 surfaces using three steel tools, collecting haptic (tool and finger accelerations), auditory (contact sounds), and force/torque applied to the surface. In \cite{fu2024touch} (SSVTP + HCT), the SSVTP dataset is expanded by incorporating HCT (Human Contact Tasks). Thus, tactile and visual data are aligned with linguistic labels. It includes training, testing, and fine-tuning splits to facilitate the development of models that interpret tactile sensations joined with visual inputs and descriptive language. Applications include open-vocabulary classification and text generation based on multimodal cues. The dataset in \cite{balasubramanian2024sens3} (SENS3) is an extensive open-access repository of multisensory data acquired from fifty surfaces during finger-surface interactions. 
The work in \cite{lima2023multimodal}, with (MMTD), focuses on dynamic tactile data for texture characterization. It was generated using a tactile-enabled finger equipped with a multi-modal tactile sensing module. This dataset incorporates data from pressure, gravity, angular rate, and magnetic field sensors, designed for texture classification.
In the paper \cite{suresh2023midastouch} they introduce (YCB-Slide). MidasTouch is a tactile perception system designed for online global localization of a vision-based touch sensor as it slides across an object’s surface. The YCB-Slide dataset, which includes real-world and simulated forceful sliding interactions between a vision-based tactile sensor and standard YCB objects, accompanies the MidasTouch framework.

The dataset in \cite{cheng2024touch100k} (Touch100k) is a large-scale touch-language-vision dataset designed for touch-centric multimodal representation. It features 100000 paired samples with tactile sensation descriptions at multiple granularities. These descriptions include sentence-level natural expressions with rich semantics, capturing contextual and dynamic relationships, as well as phrase-level descriptions highlighting key features of tactile sensations.

In \cite{dou2024tactile} (TaRF), the authors present a dataset that introduces a novel scene representation that unifies vision and touch in a shared 3D space. It estimate visual and tactile signals for specific 3D positions within a scene. Leveraging insights from multi-view geometry, they register touch signals to visual scenes and train a conditional diffusion model. This model generates tactile signals corresponding to RGB-D images rendered from neural radiance fields.

\subsection{Simulators}

Generating synthetic data to avoid the cumbersome data collection and annotation process has been commonly used in multiple domains \cite{figueira2022survey}. Similarly, simulation has been used in haptic domain to collect synthetic data. In this section we report the available simulators in the state of the art.




The simulator described in \cite{wang2022tacto} (Tacto) is designed to emulate the tactile sensations perceived by robotic systems. It offers a high degree of realism and versatility by using high-resolution vision-based tactile sensors to simulate the deformation and contact of objects with the sensor surface. Tacto provides a user-friendly interface and supports integration with popular robotics frameworks like PyBullet and ROS, facilitating seamless deployment in various robotic applications. The simulator is open-source.

In \cite{si2022taxim} (Taxim) they provide a sophisticated method to emulate the tactile responses of GelSight sensors by leveraging an example-based approach. Taxim generates realistic tactile images by mapping the deformation patterns of the sensor's gel surface to pre-recorded examples of tactile data, allowing for accurate reproduction of sensor outputs under various contact conditions. This method improves the fidelity of the simulation, making it particularly useful for machine learning and robotic manipulation tasks.

\cite{chen2023tacchi} (Tacchi) offers a streamlined and efficient approach to simulating the deformation of elastomer-based tactile sensors. Tacchi is designed to be easily integrated ("pluggable") with existing systems and achieves low computational overhead by focusing on essential deformation characteristics, thereby maintaining real-time performance. It accurately replicates the tactile feedback by simulating how the elastomer deforms upon contact with objects, ensuring that the output closely mirrors actual sensor readings. This makes Tacchi a practical tool for researchers and developers  without the burden of high computational demands.

\cite{si2024difftactile} (Difftactile) provides a sophisticated simulation environment designed to model the intricate interactions between tactile sensors and objects during robotic manipulation tasks. This simulator stands out by incorporating a differentiable physics engine, which allows for the computation of gradients with respect to the tactile interactions. This feature enables the optimization of sensor designs and control policies through gradient-based methods. Difftactile excels in simulating complex, contact-rich scenarios with high accuracy.

The application described in \cite{du2024tacipc} (Tacipc) offers an advanced simulation framework tailored for optical tactile sensors, utilizing the Finite Element Method (FEM) to model elastomer deformation. TacIPC distinguishes itself by ensuring simulations are free from geometric intersections and inversions, which enhances the stability and realism of the deformation modeling. This high-fidelity approach allows for accurate replication of how elastomer materials behave under contact, providing valuable insights for sensor design and application. 

In the paper \cite{zhao2024fots} (Fots) authors provide a rapid and realistic simulation environment for training tactile-motor skills in robots. By focusing on high-speed rendering of optical tactile feedback, Fots ensures that the simulated interactions are highly accurate and closely mimic real-world tactile responses. This simulator is specifically designed to facilitate the sim2real transfer, allowing robotic systems to effectively apply the skills learned in simulation to physical tasks. 

The paper \cite{cui2023tactile} (Tactile) offers a detailed simulation framework tailored for GelStereo sensors, which combine visual and tactile sensing capabilities. This simulator accurately replicates the tactile imprints left on the sensor's gel surface when objects make contact, capturing the intricate details of these interactions. By focusing on the visuotactile feedback, the simulation enables researchers to study and optimize the sensor's performance in a variety of tactile sensing tasks.

Finally \cite{luu2023simulation} presents a comprehensive framework for simulating, training, and deploying vision-based tactile sensors in extensive applications. This framework supports large-scale simulations that accurately model the tactile feedback from vision-based sensors, enabling the development of robust tactile sensing algorithms. By integrating learning methodologies, the framework facilitates the training of machine learning models to interpret tactile data effectively. The scalable nature of this simulation platform allows for extensive testing and optimization, making it ideal for broad deployment in various robotic and industrial contexts where tactile sensing is critical.

\section{\SecondRebuttal{Open Research Directions}} 
\label{sec:applications}

\SecondRebuttal{In this chapter, we highlight open research directions. The chapter is organized as follows: in Section \ref{sec:future_modeling}, we discuss current limitations in haptic surface modeling; in Section \ref{sec:future_processing}, we explore open research challenges in haptic surface processing; in Section \ref{sec:general_future}, we address issues related to the interoperability, implementation, and integration of current techniques; and finally, in Section \ref{sec:future_applications}, we present emerging application areas in the haptic field.}





\begin{table*}[!h]
\tiny
\centering
\resizebox{\textwidth}{!}{%
\begin{tabular}{@{}ccccccccccc@{}}
\toprule
 & \multicolumn{3}{c}{\textbf{Classification}} &  &  &  &  &  &  &  \\ \cmidrule(lr){2-4}
 &  &  &  &  &  &  &  &  &  &  \\
\multirow{-3}{*}{\textbf{Dataset}} & \multirow{-2}{*}{\textbf{\begin{tabular}[c]{@{}c@{}}Single\\ mode\end{tabular}}} & \multirow{-2}{*}{\textbf{\begin{tabular}[c]{@{}c@{}}Multi-\\ modal\end{tabular}}} & \multirow{-2}{*}{\textbf{\begin{tabular}[c]{@{}c@{}}Cross-\\ modal\end{tabular}}} & \multirow{-3}{*}{\textbf{\begin{tabular}[c]{@{}c@{}}Cross-modal\\ generation\end{tabular}}} & \multirow{-3}{*}{\textbf{Stylization}} & \multirow{-3}{*}{\textbf{\SecondRebuttal{Compression}}} & \multirow{-3}{*}{\textbf{\begin{tabular}[c]{@{}c@{}}Grasp \\ prediction\end{tabular}}} & \multirow{-3}{*}{\textbf{\begin{tabular}[c]{@{}c@{}}Contact\\ localization\end{tabular}}} & \multirow{-3}{*}{\textbf{\begin{tabular}[c]{@{}c@{}}Perceptual\\ similarity\end{tabular}}} & \multirow{-3}{*}{\textbf{\begin{tabular}[c]{@{}c@{}}Video\\ generation/\\ prediction\end{tabular}}} \\ \midrule
 &  &  &  &  &  &  &  &  &  &  \\
 &  &  &  &  &  &  &  &  &  &  \\
\multirow{-3}{*}{HaTT} & \multirow{-3}{*}{\cite{nie2021influence}} & \multirow{-3}{*}{\cite{liu2023surface}} &  & \multirow{-3}{*}{\cite{cai2022multi}} & \multirow{-3}{*}{\cite{culbertson2017ungrounded}} & \multirow{-3}{*}{\SecondRebuttal{\cite{nie2023compression}}} &  &  & \multirow{-3}{*}{\begin{tabular}[c]{@{}c@{}}\cite{vardar2019fingertip,richardson2022learning}\\ \cite{heravi2024development}\end{tabular}} &  \\
\rowcolor[HTML]{EFEFEF} 
\cellcolor[HTML]{EFEFEF} & \cellcolor[HTML]{EFEFEF} & \cellcolor[HTML]{EFEFEF} & \cellcolor[HTML]{EFEFEF} & \cellcolor[HTML]{EFEFEF} &  & \cellcolor[HTML]{EFEFEF} &  &  &  &  \\
\rowcolor[HTML]{EFEFEF} 
\cellcolor[HTML]{EFEFEF} & \cellcolor[HTML]{EFEFEF} & \cellcolor[HTML]{EFEFEF} & \cellcolor[HTML]{EFEFEF} & \cellcolor[HTML]{EFEFEF} &  & \cellcolor[HTML]{EFEFEF} &  &  &  &  \\
\rowcolor[HTML]{EFEFEF} 
\multirow{-3}{*}{\cellcolor[HTML]{EFEFEF}LMT} & \multirow{-3}{*}{\cellcolor[HTML]{EFEFEF}\textbf{\cite{strese2014haptic}}} & \multirow{-3}{*}{\cellcolor[HTML]{EFEFEF}\begin{tabular}[c]{@{}c@{}}\textbf{\cite{strese2016multimodal}}\cite{wei2022alignment}\\ \cite{zheng2016deep,zhang2021partial,zhang2021generative}\\ \cite{zhou2022multimodal,ji2015preprocessing}\end{tabular}} & \multirow{-3}{*}{\cellcolor[HTML]{EFEFEF}\begin{tabular}[c]{@{}c@{}}\cite{liu2022texture,zheng2019cross}\\ \cite{zheng2020lifelong}\end{tabular}} & \multirow{-3}{*}{\cellcolor[HTML]{EFEFEF}\begin{tabular}[c]{@{}c@{}}\cite{li2021research,cai2021visual}\\ \cite{xi2024cm,zhan2023method}\\ \cite{fang2024bidirectional}\end{tabular}} &  & \multirow{-3}{*}{\cellcolor[HTML]{EFEFEF}\begin{tabular}[c]{@{}c@{}}\SecondRebuttal{\cite{hassen2020pvc, zhu2022perceptual}}\\ \SecondRebuttal{\cite{liu2023online}}\end{tabular}} &  &  &  &  \\
 &  &  &  &  &  &  &  &  &  &  \\
 &  &  &  &  &  &  &  &  &  &  \\
\multirow{-3}{*}{FoS} &  &  &  &  &  &  & \multirow{-3}{*}{\begin{tabular}[c]{@{}c@{}}\textbf{\cite{calandra2017feeling}}\cite{yang2024binding}\\ \cite{cui2020self,liu2023deep}\\ \cite{liu2023self,zhang2023grasp}\end{tabular}} &  &  &  \\
\rowcolor[HTML]{EFEFEF} 
\cellcolor[HTML]{EFEFEF} & \cellcolor[HTML]{EFEFEF} & \cellcolor[HTML]{EFEFEF} &  & \cellcolor[HTML]{EFEFEF} &  &  &  &  &  &  \\
\rowcolor[HTML]{EFEFEF} 
\cellcolor[HTML]{EFEFEF} & \cellcolor[HTML]{EFEFEF} & \cellcolor[HTML]{EFEFEF} &  & \cellcolor[HTML]{EFEFEF} &  &  &  &  &  &  \\
\rowcolor[HTML]{EFEFEF} 
\multirow{-3}{*}{\cellcolor[HTML]{EFEFEF}Fabrics} & \multirow{-3}{*}{\cellcolor[HTML]{EFEFEF}\cite{wang2019fabric,cao2023learn}} & \multirow{-3}{*}{\cellcolor[HTML]{EFEFEF}\cite{wei2022alignment,zhang2021generative}} &  & \multirow{-3}{*}{\cellcolor[HTML]{EFEFEF}} &  &  &  &  &  &  \\
 &  &  &  &  &  &  &  &  &  &  \\
 &  &  &  &  &  &  &  &  &  &  \\
\multirow{-3}{*}{Texplorer} & \multirow{-3}{*}{\textbf{\cite{strese2019haptic}}} & \multirow{-3}{*}{\begin{tabular}[c]{@{}c@{}}\textbf{\cite{strese2017content,strese2019haptic}}\\ \cite{khojasteh2023multimodal,liu2018surface}\\ \cite{liu2023surface,wei2021multimodal}\end{tabular}} & \multirow{-3}{*}{\cite{zheng2020lifelong}} & \multirow{-3}{*}{\cite{song2023cross}} &  &  &  &  & \multirow{-3}{*}{\cite{priyadarshini2019perceptnet}} &  \\
\rowcolor[HTML]{EFEFEF} 
\cellcolor[HTML]{EFEFEF} & \cellcolor[HTML]{EFEFEF} & \cellcolor[HTML]{EFEFEF} & \cellcolor[HTML]{EFEFEF} & \cellcolor[HTML]{EFEFEF} &  &  &  &  &  &  \\
\rowcolor[HTML]{EFEFEF} 
\cellcolor[HTML]{EFEFEF} & \cellcolor[HTML]{EFEFEF} & \cellcolor[HTML]{EFEFEF} & \cellcolor[HTML]{EFEFEF} & \cellcolor[HTML]{EFEFEF} &  &  &  &  &  &  \\
\rowcolor[HTML]{EFEFEF} 
\multirow{-3}{*}{\cellcolor[HTML]{EFEFEF}ViTac} & \multirow{-3}{*}{\cellcolor[HTML]{EFEFEF}\begin{tabular}[c]{@{}c@{}}\textbf{\cite{luo2018vitac}}\\ \cite{cao2020spatio,rouhafzay2020transfer}\end{tabular}} & \multirow{-3}{*}{\cellcolor[HTML]{EFEFEF}\cite{rouhafzay2020transfer}} & \multirow{-3}{*}{\cellcolor[HTML]{EFEFEF}\textbf{\cite{luo2018vitac}}\cite{rouhafzay2020transfer}} & \multirow{-3}{*}{\cellcolor[HTML]{EFEFEF}\cite{lee2019touching}} &  &  &  &  &  &  \\
 &  &  &  &  &  &  &  &  &  &  \\
 &  &  &  &  &  &  &  &  &  &  \\
\multirow{-3}{*}{MTaF} &  &  &  &  &  &  & \multirow{-3}{*}{\textbf{\cite{calandra2018more}}\cite{dave2024multimodal}} &  &  &  \\
\rowcolor[HTML]{EFEFEF} 
\cellcolor[HTML]{EFEFEF} &  &  &  & \cellcolor[HTML]{EFEFEF} &  & \cellcolor[HTML]{EFEFEF} &  &  &  &  \\
\rowcolor[HTML]{EFEFEF} 
\cellcolor[HTML]{EFEFEF} &  &  &  & \cellcolor[HTML]{EFEFEF} &  & \cellcolor[HTML]{EFEFEF} &  &  &  &  \\
\rowcolor[HTML]{EFEFEF} 
\multirow{-3}{*}{\cellcolor[HTML]{EFEFEF}Haptex} &  &  &  & \multirow{-3}{*}{\cellcolor[HTML]{EFEFEF}\cite{cai2022gan}} &  & \multirow{-3}{*}{\cellcolor[HTML]{EFEFEF}\SecondRebuttal{\cite{liu2023online}}} &  &  &  &  \\
 &  &  &  &  &  &  &  &  &  &  \\
 &  &  &  &  &  &  &  &  &  &  \\
\multirow{-3}{*}{VisGel} &  &  & \multirow{-3}{*}{\cite{li2023learning}} & \multirow{-3}{*}{\cite{yang2023generating}} &  &  &  &  &  & \multirow{-3}{*}{\cite{li2023learning}} \\
\rowcolor[HTML]{EFEFEF} 
\cellcolor[HTML]{EFEFEF} &  & \cellcolor[HTML]{EFEFEF} & \cellcolor[HTML]{EFEFEF} & \cellcolor[HTML]{EFEFEF} &  &  & \cellcolor[HTML]{EFEFEF} & \cellcolor[HTML]{EFEFEF} &  & \cellcolor[HTML]{EFEFEF} \\
\rowcolor[HTML]{EFEFEF} 
\cellcolor[HTML]{EFEFEF} &  & \cellcolor[HTML]{EFEFEF} & \cellcolor[HTML]{EFEFEF} & \cellcolor[HTML]{EFEFEF} &  &  & \cellcolor[HTML]{EFEFEF} & \cellcolor[HTML]{EFEFEF} &  & \cellcolor[HTML]{EFEFEF} \\
\rowcolor[HTML]{EFEFEF} 
\multirow{-3}{*}{\cellcolor[HTML]{EFEFEF}ObjectFolder} &  & \multirow{-3}{*}{\cellcolor[HTML]{EFEFEF}\textbf{\cite{gao2023objectfolder}}\cite{li2023vito}} & \multirow{-3}{*}{\cellcolor[HTML]{EFEFEF}\begin{tabular}[c]{@{}c@{}}\textbf{\cite{gao2021objectfolder}}\cite{gao2023objectfolder}\\ \cite{li2023learning,yang2024binding}\end{tabular}} & \multirow{-3}{*}{\cellcolor[HTML]{EFEFEF}\textbf{\cite{gao2023objectfolder}}} &  &  & \multirow{-3}{*}{\cellcolor[HTML]{EFEFEF}\textbf{\cite{gao2023objectfolder}}\cite{yang2024binding}} & \multirow{-3}{*}{\cellcolor[HTML]{EFEFEF}\textbf{\cite{gao2023objectfolder}}} &  & \multirow{-3}{*}{\cellcolor[HTML]{EFEFEF}\cite{li2023learning}} \\
 &  &  &  &  &  &  &  &  &  &  \\
 &  &  &  &  &  &  &  &  &  &  \\
\multirow{-3}{*}{Touch and Go} & \multirow{-3}{*}{\begin{tabular}[c]{@{}c@{}}\cite{dave2024multimodal,lei2024vit}\\ \cite{yang2024binding}\end{tabular}} & \multirow{-3}{*}{\begin{tabular}[c]{@{}c@{}}\textbf{\cite{yang2022touch}}\cite{dave2024multimodal}\\ \cite{lei2024vit,lyu2024omnibind}\end{tabular}} & \multirow{-3}{*}{\cite{yang2024binding,li2023learning}} & \multirow{-3}{*}{\begin{tabular}[c]{@{}c@{}}\cite{yang2023generating,yang2024binding}\\ \cite{fang2024bidirectional}\end{tabular}} & \multirow{-3}{*}{\begin{tabular}[c]{@{}c@{}}\textbf{\cite{yang2022touch}}\\ \cite{yang2024binding,yang2023generating}\end{tabular}} &  & \multirow{-3}{*}{\textbf{\cite{yang2022touch}}\cite{yang2024binding}} &  &  & \multirow{-3}{*}{\textbf{\cite{yang2022touch}}\cite{li2023learning}} \\
\rowcolor[HTML]{EFEFEF} 
\cellcolor[HTML]{EFEFEF} &  &  &  &  &  &  & \cellcolor[HTML]{EFEFEF} &  &  &  \\
\rowcolor[HTML]{EFEFEF} 
\cellcolor[HTML]{EFEFEF} &  &  &  &  &  &  & \cellcolor[HTML]{EFEFEF} &  &  &  \\
\rowcolor[HTML]{EFEFEF} 
\multirow{-3}{*}{\cellcolor[HTML]{EFEFEF}PoseIt} &  &  &  &  &  &  & \multirow{-3}{*}{\cellcolor[HTML]{EFEFEF}\textbf{\cite{kanitkar2022poseit}}} &  &  &  \\
 &  &  &  &  &  &  &  &  &  &  \\
 &  &  &  &  &  &  &  &  &  &  \\
\multirow{-3}{*}{SSVTP} & \multirow{-3}{*}{\cite{yang2024binding}} & \multirow{-3}{*}{\textbf{\cite{kerr2022self}}} & \multirow{-3}{*}{\cite{yang2024binding}} &  &  &  & \multirow{-3}{*}{\cite{yang2024binding}} & \multirow{-3}{*}{\textbf{\cite{kerr2022self}}} &  &  \\
\rowcolor[HTML]{EFEFEF} 
\cellcolor[HTML]{EFEFEF} &  & \cellcolor[HTML]{EFEFEF} &  &  &  &  &  &  &  &  \\
\rowcolor[HTML]{EFEFEF} 
\cellcolor[HTML]{EFEFEF} &  & \cellcolor[HTML]{EFEFEF} &  &  &  &  &  &  &  &  \\
\rowcolor[HTML]{EFEFEF} 
\multirow{-3}{*}{\cellcolor[HTML]{EFEFEF}MPI-10} &  & \multirow{-3}{*}{\cellcolor[HTML]{EFEFEF}\textbf{\cite{khojasteh2024robust}}} &  &  &  &  &  &  &  &  \\
 &  &  &  &  &  &  &  &  &  &  \\
 &  &  &  &  &  &  &  &  &  &  \\
\multirow{-3}{*}{SSVTP + HCT} &  & \multirow{-3}{*}{\textbf{\cite{fu2024touch}}} & \multirow{-3}{*}{\textbf{\cite{fu2024touch}}} &  &  &  &  &  &  &  \\
\rowcolor[HTML]{EFEFEF} 
\cellcolor[HTML]{EFEFEF} &  & \cellcolor[HTML]{EFEFEF} &  & \cellcolor[HTML]{EFEFEF} &  &  &  & \cellcolor[HTML]{EFEFEF} &  &  \\
\rowcolor[HTML]{EFEFEF} 
\cellcolor[HTML]{EFEFEF} &  & \cellcolor[HTML]{EFEFEF} &  & \cellcolor[HTML]{EFEFEF} &  &  &  & \cellcolor[HTML]{EFEFEF} &  &  \\
\rowcolor[HTML]{EFEFEF} 
\multirow{-3}{*}{\cellcolor[HTML]{EFEFEF}TaRF} &  & \multirow{-3}{*}{\cellcolor[HTML]{EFEFEF}\textbf{\cite{dou2024tactile}}} &  & \multirow{-3}{*}{\cellcolor[HTML]{EFEFEF}\textbf{\cite{dou2024tactile}}} &  &  &  & \multirow{-3}{*}{\cellcolor[HTML]{EFEFEF}\textbf{\cite{dou2024tactile}}} &  &  \\ \bottomrule
\end{tabular}%
}
\caption{This table illustrates the tasks performed on the most used datasets presented in Table \ref{tab:datasets}, highlighting in \textbf{bold} the baselines methods.}
\label{tab:tasks}
\end{table*}

\subsection{\textbf{Haptic surface modeling}} 
\label{sec:future_modeling}

\Rebuttal{In haptic surface modeling, available datasets and modeling methods leave two main research questions open.}
\\
\\

\Rebuttal{\textit{What is the best way to collect highly precise haptic data?} As shown in Chapter \ref{sec:haptic_surface_modeling}, the literature remains divided on the optimal representation and collection of haptic data. On one hand, vision-based data collection—such as images and videos—has gained significant traction in recent years due to its ability to capture spatially distributed surface data and the wide availability of advanced processing techniques in this domain. However, a precise, high-resolution mapping between RGB images and specific spatial points remains elusive, leading to difficulties in estimating tactile features like friction and stiffness. On the other hand, parametric signals can capture highly precise data but struggle to generalize knowledge to neighboring points on a surface. Recent approaches \cite{shin2018geometry,shin2020hybrid,heravi2020learning} have attempted to combine the strengths of both methods, but their applicability is limited by the lack of available datasets. As discussed in Chapter \ref{sec:datasets}, the current literature lacks comprehensive datasets that provide matched haptic maps and one-dimensional signals, which constrains further research developments.}

\Rebuttal{\textit{How to validate the modeling quantitatively?} Researchers have primarily validated their models through user studies, which, while effective, are difficult and costly to replicate. Another approach involves using the model for downstream tasks like material classification, but this method validates the collected features rather than the model itself. These limitations arise from the inherent challenge of ensuring that the same physical objects are available to all practitioners.
One potential solution is to rely on digitally stored 3D models \cite{gao2021objectfolder}. However, these models are often designed either as implicit representations, which only allow the collection of specific signals, or rely on 3D engines that lack tools for incorporating tactile samples into scenes. The availability of a standardized set of shared 3D models, complete with ground-truth data, would enable researchers to quantitatively evaluate and compare different modeling techniques. This would foster the development of a unified validation framework and comparison metrics, ultimately advancing the field.}

\subsection{\textbf{Haptic surface processing}} 
\label{sec:future_processing}

\Rebuttal{In this chapter we provide an entry point for practitioners on the main tasks currently tackled in the haptic signal processing domain and highlight current research directions.}
Starting from the datasets listed in Section \ref{sec:datasets}, in Tab. \ref{tab:tasks} we provide a thorough classification of the available works for each task. While some of them, like classification and compression, are akin to other domains and straightforward to understand, other tasks are very specific to \Rebuttal{the haptic domain}, like contact localization and grasp prediction. \Rebuttal{As shown in the previous paragraph, processing techniques are deeply intertwined with modeling, as they are used as downstream tasks to validate the data modeling, or, vice-versa, they provide strict requirements about the information that should be contained in the data.}

For a comprehensive overview, in the next paragraphs we delve into each task, highlighting their peculiarities.

\paragraph{\textbf{Classification}} It is one of the tasks mainly performed when dealing with haptic data. As the collected datasets usually provide ground truths in different domains (e.g. haptic maps, RGB images, audio produced by the object upon contact) for the same sample, classification can be performed according to different strategies: \textit{single mode}, \textit{multi-modal}, and \textit{cross-modal}. Works adopting the \textit{single mode} strategy aim to classify materials/objects by training and testing in the same domain. For approaches employing only haptic data, this task is usually performed on tactile maps \cite{luo2018vitac, strese2019haptic, wang2019fabric, cao2020spatio, rouhafzay2020transfer, gao2021objectfolder, wei2021multimodal, cao2023learn, gao2023objectfolder, dave2024multimodal, lei2024vit, yang2024binding}, since dealing with images enables the usage of established approaches, such as deep neural networks (DNNs). Differently, other works, such as \cite{strese2014haptic, nie2021influence} proposed a material classification method that exploits acceleration data, achieving state-of-the-art results and showing that acceleration data establish a unique fingerprint for each material. \textit{Multi-modal} approaches employ multiple data sources to classify materials and objects. These works rely mainly on visual and haptic data \cite{zheng2016deep, rouhafzay2020transfer, zhang2021generative, zhang2021partial, kerr2022self, wei2022alignment, yang2022touch, zhou2022multimodal, li2023vito, liu2023surface, dave2024multimodal, dou2024tactile, fu2024touch, lyu2024omnibind, yang2024binding}, since their combination allows for the exploitation of both mono-dimensional and multi-dimensional data. However, also audio cues are widely used jointly with haptic data \cite{ji2015preprocessing, strese2016multimodal, strese2017content, liu2018surface, strese2019haptic, gao2021objectfolder, wei2021multimodal, gao2023objectfolder, khojasteh2023multimodal, khojasteh2024robust, lei2024vit} enabling approaches that leverage the combination of all these data types. \textit{Cross-modal} classification is a task that leverages knowledge acquired on one data domain to classify data from another one. This allows, for example, to exploit data domains that are more easily accessible, such as images and audio domains, and apply knowledge on an unexplored domain, like haptic. Cross-modal classification has been performed with audio data in \cite{liu2022texture}, and using visual data in \cite{luo2018vitac, rouhafzay2020transfer, liu2022texture, fu2024touch}. Furthermore, a subset task of the cross-modal classification application is \textit{cross-modal retrieval} \cite{zheng2019cross, zheng2020lifelong, gao2021objectfolder, gao2023objectfolder, li2023learning}. Cross-modal retrieval sits between classification and generation applications. The core idea consists of retrieving the data sample in one domain based on a query from another domain by finding the best correspondence within the dataset used to train the system.


\paragraph{\textbf{Cross-modal generation}} It refers to the creation of content in one sensory modality based on the information gathered from another one. This process involves translating the haptic information of a surface across different types of media. Recently, this task has gained popularity in many real-world applications, since it allows to collect data in one modality and to transfer the information to different ones. From the haptic perspective, cross-modal generation can be applied to generate vibro-tactile acceleration signals from other types of data, such as texture images \cite{cai2021visual, li2021research, cai2022gan, cai2022multi, fang2024bidirectional, xi2024cm}. Another approach involves mapping acoustic and acceleration signals, exploiting their similarities \cite{gao2023objectfolder, zhan2023method}. Furthermore, visual and acoustic domains can also be combined \cite{song2023cross}. Another data type widely employed in literature for the cross-modal generation task is the haptic map, which enables the usage of computer vision algorithms to map RGB images into haptic data \cite{lee2019touching, gao2023objectfolder, yang2023generating, dou2024tactile, fang2024bidirectional, yang2024binding}. 

\paragraph{\textbf{Stylization}} It is a recent application that, given a desired haptic map, aims at "stylizing" the RGB image of an object, modifying its appearance to reflect the tactile characteristics accurately. This application has been applied mainly to the Touch And Go dataset using methods based on contrastive learning \cite{yang2022touch} or diffusion models \cite{yang2023generating, yang2024binding}. \Rebuttal{A method exploiting height maps instead of haptic maps is presented in \cite{tymms2020appearance}. Here the authors define an optimization process that aims at changing the roughness of an object while preserving its visual appearance.} Despite this task emerging from the use of RGB and Haptic Maps data, the HaTT dataset has also been previously employed. In \cite{culbertson2017ungrounded}, the authors exploited a physical surface as a constraint, modifying the accelerations emitted by the haptic device to render a different surface according to the model collected in HaTT. \Rebuttal{A different approach employing a perceptual metric is presented in \cite{piovarvci2020fabrication}. Here, the authors use a data-driven method to properly tune the haptic feedback given by digital drawing tools (i.e., drawing surface and stylus), defining different drawing styles.}

\SecondRebuttal{\paragraph{\textbf{Compression}} It is a key task in the field of signal processing, with the goal of enabling efficient transmission of haptic data through compression and reconstruction strategies. This is particularly important in scenarios requiring real-time communication, such as in medical applications. Research in this area has primarily focused on vibro-tactile signals and has explored both traditional signal processing techniques and deep neural network (DNN)-based approaches. In \cite{hassen2020pvc} they proposed a vibro-tactile perceptual coding method that utilizes a sparse linear prediction strategy. To validate the reconstruction efficiency objectively, the authors exploited the \textit{ST-SIM} metric, previously introduced in \cite{hassen2019vibrotactile, hassen2019subjective}. This approach has been also integrated into the tactile codec standard
by the IEEE1918.1.1 Haptic Codecs Task Group for the Tactile
Internet. In another work \cite{nie2023compression}, they convert haptic signals into grayscale images, which are then compressed using conventional still image compression techniques. On the DNN side, in \cite{zhu2022perceptual} they introduced a transformer-based architecture for perceptual compression of haptic data, while in \cite{liu2023online} they combined stacked autoencoders with gated recurrent units for online compression.}

\paragraph{\textbf{Grasp prediction}} This is a key task in robotics, to instruct a robotic hand to manipulate an object, performing tasks like moving or positioning it. While not being strictly modeling-related, it requires the understanding of the stiffness and friction properties of an object, thus modeling of haptic properties plays a fundamental role in this task. These algorithms rely mainly on vision-based sensors to gather haptic information on the part of the object occluded by robotic fingers that cannot be seen by external cameras. To perform grasp prediction various strategies have been developed. In \cite{calandra2017feeling, gao2023objectfolder}, the authors rely on the difference between the initial and final grasp position to train a DNN to predict the success of the grasp. 
Furthermore, in \cite{calandra2018more}, the authors introduced a conditioning factor that enables the DNN to adjust the grasping action dynamically if the first attempt fails. Starting from the same idea of training the classifier to predict if the grasp will be successful or not, in \cite{cui2020self, liu2023deep} the authors designed a self-attention mechanism, while in \cite{yang2022touch, liu2023self, zhang2023grasp, dave2024multimodal, yang2024binding} contrastive learning frameworks have been adopted. All these works employ only tactile maps and RGB images to perform grasp prediction. The work in \cite{kanitkar2022poseit} represents the only exception, where, to understand if grasping will be successful in challenging conditions, such as while shaking the object, the authors also employed mono-dimensional signals collected using force/torque sensors to train the classifier. \Rebuttal{A different approach is taken by the method of \cite{lau2016tactile}, in which the authors employ a prior knowledge of the mesh of an object to identify the salient points at which a successful grasp can be made.} Furthermore, grasp prediction is of particular interest for the works that aim at bridging the gap between simulation and reality, as presented in \cite{gao2023objectfolder, zhang2023grasp}. \Rebuttal{An alternative approach is proposed in \cite{sundaram2019learning}, where the authors developed a dataset of user grasps recorded using a sensor-equipped glove. A neural network was then used for object identification, weight prediction, hand pose estimation, and analyzing hand region correlations during grasping.}


\paragraph{\textbf{Contact localization}} It is another key application in robotics. This application aims at identifying the exact point on an object where a sample has been taken. Initially, the primary goal of this task was to determine where an object was touched to teach robots how to grasp object correctly \cite{kerr2022self, suresh2023midastouch}. Recently, however, research has moved towards identifying the location on an RGB image or a 3D model where other types of data, such as acoustic signals \cite{gao2023objectfolder} or haptic signals \cite{dou2024tactile}, were generated. 


\paragraph{\textbf{Perceptual similarity}} It is an application aiming at assessing how human touch feels different textures or surfaces. The goal of this task is to identify models capable of predicting human perception based on physical interactions with surfaces. In \cite{vardar2019fingertip} and \cite{richardson2022learning}, the authors propose an experiment using the HaTT dataset where users has to feel different textures by exploring randomly real surfaces made of different materials. The goal of these works is to identify the similarities that different textures present to cluster them according to the feature extracted. In \cite{priyadarshini2019perceptnet}, the authors try to cluster the textures, according to the feature perceived by user, into a feature space. To do this they employ respectively a K-means strategy and a DNN method. In \cite{heravi2024development}, the authors exploit GelSight images to capture textures information and, by modeling the user interaction with them, they create virtual textures that close up with their real counterpart.


\paragraph{\textbf{Video generation and prediction}} It is a very specific application that relies on video streams from vision-based sensors like GelSight or standard RGB cameras. The goal of this task is to retrieve missing frames within videos collected from a specific data source. In \cite{yang2022touch}, two approaches are explored: single-mode, which relies only on previous frames to generate missing ones, and multimodal, which leverages previous frames coming from different data sources. Differently, in \cite{li2023learning}, the authors rely on a cross-generation framework, using GelSight-captured video information to generate missing frames in RGB videos and vice-versa.


\bigskip

\SecondRebuttal{This chapter has explored the diverse application domains where haptic datasets have been utilized. The tasks analyzed highlight the need for an interconnected and multimodal research approach to address open challenges in the field. Building on this foundation, major research efforts should focus on improving the realism and adaptability of haptic models to better capture fine-grained textures and dynamic properties of real-world surfaces. Current methods are typically based on predefined sets of materials, but future work should prioritize adaptive frameworks that can interpret scenario information in real time so that they can generalize for any type of surface. This shift introduces not only a data collection problem, but also a scalability challenge, as many existing architectures struggle to run efficiently in runtime environments, posing limitations for robotic and extended reality applications. Finally, as haptic research advances, establishing reliable, objective evaluation metrics becomes crucial. While defining such metrics may be more straightforward in robot-centric applications, scenarios involving human-in-the-loop present additional complexities. The inherent bias in subjective evaluations can undermine standardization efforts, making it difficult to develop universally accepted assessment frameworks.}

\subsection{\SecondRebuttal{\textbf{Interoperability, implementation and integration}}}
\label{sec:general_future}

\SecondRebuttal{Beyond specific technical challenges, haptics research is increasingly influenced by a number of issues that cut across domains and applications. This section outlines the main limitations in standardisation, multimodal integration, real-time capabilities and evaluation practices that need to be addressed to advance the field.}

\SecondRebuttal{\textit{Missing Standardization.} As discussed in Section~\ref{sec:processing_pipeline}, one of the major limitations hindering cohesive and scalable research in haptics is the absence of a standardized processing pipeline. This issue is twofold: on the one hand, there is no universally accepted data type or format that has proven clearly superior in encoding haptic information. On the other hand, existing datasets lack the scale (in terms of sample size) and diversity (in terms of class variety) that have led to breakthroughs in other domains, such as ImageNet in computer vision. This not only limits the range of features and representations that can be explored and benchmarked, but also limits the effectiveness of data-driven and machine learning frameworks.}

\SecondRebuttal{\textit{Multimodality.} The integration of multiple data modalities has been a driving force in areas like vision and language. In haptics, multimodality can be both intra-domain and cross-domain. Intra-domain multimodality currently lacks a unified framework to relate different haptic data types—such as haptic maps and mono-dimensional signals—and to bridge human- and robot-based sensing. Enabling seamless conversion between data formats would allow for dataset expansion and deeper convergence across representations. Cross-domain multimodality remains an underexplored avenue: integrating haptic signals with visual and textual information (e.g., in large language or vision-language models) could dramatically increase the availability of training data and enable richer semantic understanding.}

\SecondRebuttal{\textit{Real-Time Performance.} At present, only mono-dimensional haptic signals are effectively used in real-time applications such as teleoperation, due to their relatively low data complexity. More complex data representations—such as haptic images or 3D tactile reconstructions—are still far from being captured, processed, and rendered in real time. Future research should focus on improving computational efficiency and designing hardware capable of low-latency acquisition and feedback.}


\SecondRebuttal{\textit{Evaluation and User Experience.} There is a pressing need for better evaluation methodologies in haptic research. New metrics should be developed to assess the fidelity, expressiveness, and perceptual quality of haptic modeling algorithms. While physical modeling can use metrics like MSE or L1 distance, these fail in evaluating results based on perception, which is subjective for definition. Promising efforts, such as the metric proposed in \cite{hassen2019vibrotactile, hassen2019subjective}, combine objective computation with subjective validation, but this approach is still underdeveloped. To advance the field, new perceptually grounded metrics and standardized testing protocols across devices and contexts are essential for ensuring reproducibility and improving user experience, especially in virtual and remote interactions applications.}


\subsection{\textbf{Applications}}
\label{sec:future_applications}

\Rebuttal{Future research in the field of haptics will also encompass a wide range of application-driven studies, where the final use cases are well-defined. In this section, we provide a brief summary of such applications, specifically focusing on the fields of extended reality (XR) and medicine. Comprehensive reviews of current applications of haptic technologies can be found in \cite{giri2021application, zhu2022soft, pacheco2024haptic}.}
\\


\Rebuttal{\textit{Extended Reality.} A significant research gap in haptic technologies is the integration of haptic data representations into graphic 3D engine pipelines, as highlighted in studies such as \cite{wakita2008texturebased, kamuro2012haptic, kim2011haptic, costes2018haptic}. Current technologies often rely on custom libraries that are challenging to access and manipulate, with no shared open-source library available—a resource that would significantly benefit researchers. This issue is likely tied to the aforementioned lack of a standardized pipeline and it represents a fundamental open research direction.}

\Rebuttal{\textit{Training.} Haptic technology is increasingly employed to train expert workers who must operate in hazardous conditions that are challenging to replicate safely. It allows them to practice procedures and techniques in a simulated environment, improving their skills and reducing the risk of errors during real-life procedures.}


\Rebuttal{\textit{Telemedicine.} Telemedicine leverages information and telecommunication technology to deliver medical services remotely, overcoming geographical barriers between physicians and patients. While telemedicine has largely focused on communication-related aspects \cite{alenoghena2023telemedicine}, integrating haptics enhances remote medical examinations by transmitting touch sensations like pressure and vibration. One key aspect to enhance this practice consists in improving the quality of the signal modeled and transmitted. For example, a haptic device attached to a surgeon's gloves can simulate tactile feedback during a robot-assisted surgery, allowing the surgeon to feel tissue resistance and pressure, thus improving the quality of care.}

\Rebuttal{\textit{Rehabilitation.} Haptic technology is employed in applications like prosthetics to provide a more natural and intuitive sense of touch. This enables amputees to achieve greater control and enhanced dexterity, leading to increased independence and an improved quality of life. In this context, developing novel and efficient methods for modeling haptic data to enable real-time haptic feedback is crucial for supporting more effective rehabilitation.}

\Rebuttal{\textit{Surgery.} In minimally invasive and robotic surgeries, haptic technology provides surgeons with an intuitive sense of touch, enabling them to perform procedures with greater precision and control. This can lead to improved surgical outcomes, shorter recovery times, and reduced risk of complications.}

\Rebuttal{\textit{Communication.} Although an initial protocol for transmitting haptic data has been developed \cite{steinbach2018haptic}, optimizing the compression and reconstruction of haptic data remains a critical challenge for various applications (e.g. telemedicine). In this context, as highlighted in \cite{emami2024survey} research remains open and require further exploration.}

\section{Conclusions}
\label{sec:conclusions}

Digitalizing real objects and surfaces to enable virtual interaction requires following a rather strict pipeline, which stems from the definition of the sensing device and ends with the user perception study. 

In this paper we have decided to delve into one of the mostly unexplored areas in haptics, namely the representation and modeling of surfaces. To do so we have identified the most salient properties of objects, providing an exhaustive overview of the existing datasets and the related applications. 

Our analysis has shown how the non-standardisation of haptic data modeling provides an open research field, where a common standard has yet to emerge. Consequently, the literature is still rapidly evolving opening up to new research challenges in the domain of haptics.

\section*{CRediT authorship contribution statement}
\textbf{Antonio Luigi Stefani}: Conceptualization, Investigation, Visualization, Writing – original draft, Writing – review \& editing. \textbf{Niccolò Bisagno}: Conceptualization, Investigation, Writing – original draft, Writing – review \& editing. \textbf{Andrea Rosani}: Investigation, Writing – original draft, Writing – review \& editing. \textbf{Nicola Conci}: Conceptualization, Project administration, Supervision, Writing – review \& editing. \textbf{Francesco De Natale}: Funding acquisition, Supervision, Writing – review \& editing.

\section*{Data availability}
No data have been used to this project.

\section*{Acknowledgments}
We acknowledge the support of the MUR PNRR project iNEST-Interconnected Nord-Est Innovation Ecosystem (ECS00000043) funded by the European Union under NextGenerationEU. Views and opinions expressed are however those of the author(s) only and do not necessarily reflect those of the European Union or The European Research Executive Agency. Neither the European Union nor the granting authority can be held responsible for them.

\bibliographystyle{elsarticle/elsarticle-num} 
\bibliography{refs}






\end{document}